\documentclass[10pt,journal,twocolumn,twoside]{IEEEtran}
\usepackage{optional}

\usepackage[cmex10]{amsmath}
\usepackage{amssymb}

\usepackage{array}
\usepackage[caption=false,font=footnotesize]{subfig}
\usepackage{fixltx2e}
\usepackage{url} 

\usepackage{graphicx}
\usepackage{multirow}
\usepackage[noadjust]{cite}
\usepackage{epstopdf}

\newcommand\blfootnote[1]{%
  \begingroup
  \renewcommand\thefootnote{}\footnote{#1}%
  \addtocounter{footnote}{-1}%
  \endgroup
}

\renewcommand{\vec}{\mathbf}
\newcommand{\greekvec}{\boldsymbol}

\newcommand{\mat}{\mathbf}

\newcommand{\argmax}{\operatornamewithlimits{arg\ max}}

\begin{document}
  \title{Identifying Cover Songs Using Information-Theoretic Measures of Similarity\opt{preprint}{\blfootnote{This work has been submitted to the IEEE for possible publication. Copyright may be transferred without notice, after which this version may no longer be accessible.}}}
  \author{
	  Peter~Foster\opt{journal}{,~\IEEEmembership{Student~Member,~IEEE,}}\opt{preprint}{,}
	  Simon~Dixon,
	  and~Anssi~Klapuri
      \thanks{Copyright~\copyright~2015 IEEE. Personal use of this material is permitted.
      However, permission to use this material for any other purposes must be
      obtained from the IEEE by sending a request to pubs-permissions@ieee.org.}
	  \thanks{P.F.~is funded by an Engineering and Physical Sciences Research Council Doctoral Training Account studentship.}
      \thanks{P.F.~and S.D.~are with the School of Electronic Engineering and Computer Science, Queen Mary University of London, London, E1~4NS, UK. A.K.~is with Tampere University of Technology, Tampere, Finland, FI-33720. (Email: peter.foster@eecs.qmul.ac.uk; simon.dixon@eecs.qmul.ac.uk; anssi.klapuri@tut.fi)}}

\markboth{IEEE/ACM Transactions on Audio, Speech and Language Processing, VOL.~23, NO.~6, June 2015}%
{Foster \MakeLowercase{\textit{et al.}}: Identifying Cover Songs Using Information-Theoretic Measures of Similarity}

  \maketitle

\begin{abstract}
This paper investigates methods for quantifying similarity between audio signals, specifically for the task of of cover song detection. We consider an information-theoretic approach, where we compute pairwise measures of predictability between time series. We compare discrete-valued approaches operating on quantised audio features, to continuous-valued approaches. In the discrete case, we propose a method for computing the normalised compression distance, where we account for correlation between time series. In the continuous case, we propose to compute information-based measures of similarity as statistics of the prediction error between time series. We evaluate our methods on two cover song identification tasks using a data set comprised of 300 Jazz standards and using the Million Song Dataset. For both datasets, we observe that continuous-valued approaches outperform discrete-valued approaches. We consider approaches to estimating the normalised compression distance (NCD) based on string compression and prediction, where we observe that our proposed normalised compression distance with alignment (NCDA) improves average performance over NCD, for sequential compression algorithms. Finally, we demonstrate that continuous-valued distances may be combined to improve performance with respect to baseline approaches. Using a large-scale filter-and-refine approach, we demonstrate state-of-the-art performance for cover song identification using the Million Song Dataset.
\end{abstract}

\opt{journal}{
  
  \ifCLASSOPTIONpeerreview
   \begin{center} \bfseries EDICS Category: AUD-CONT \end{center}
  \fi
  \IEEEpeerreviewmaketitle
}


\section{Introduction}
\opt{journal}{\IEEEPARstart{I}{n}}\opt{preprint}{In} the field of music content analysis, quantifying similarity between audio signals has received a substantial amount of interest \cite{casey2008content}. Owing to the proliferation of music in digital formats, there exists potential for applications using music similarity techniques, in a wide range of domains. At the level of individual tracks, these domains span audio fingerprinting \cite{cano2005review}, cover song identification \cite{serra2011identification}, artist identification \cite{fujihara2010modeling,mandel2005song} and genre classification \cite{scaringella2006automatic}. Applications can be distinguished according to their \textit{degree of specificity} \cite{casey2008content}, referring to the level of granularity required for retrieving audio tracks from a collection, given a query track. For example, in audio fingerprinting, the required specificity is high, since the set of possible tracks corresponding to a particular recording is typically small, in relation to the data set. In contrast, genre classification requires low specificity, since the set of tracks sharing a common genre is potentially large, in relation to the data set.

A cover song may be defined as a rendition of a previously recorded piece of music \cite{serra2008chroma}. Cover song identification is deemed to have mid-level, diffuse specificity, since cover songs may differ from the original song in various musical facets, including rhythm, tempo, melody, harmonisation, instrumentation, lyrics and musical form. Correspondingly, cover song identification remains a challenging problem \cite{serra2011identification}.

In this work, we investigate methods for cover song identification that are based on quantifying pairwise predictability between sequences. From a music-psychological perspective, the significance of intrinsic predictability in musical sequences has been reflected on by Meyer \cite{meyer1956music}, who considers the possibility of using Shannon's information theory \cite{shannon1949mathematical} to quantify predictive uncertainty. Statistical learning is implicated in forming musical expectations \cite{huron2006sweet}; a successful approach to modelling expectations in response to an unfolding stream of musical events involves estimating sequential statistical models and computing information-theoretic measures of predictive uncertainty \cite{pearce2012auditory}. As exemplified in \cite{abdallah2009information}, an information-theoretic approach admits a rich conceptual framework for quantifying predictive uncertainty in musical sequences. For our own purposes in cover song identification, we seek to establish if an information-theoretic approach might be useful for determining pairwise similarity between tracks.

Based on our previous work \cite{foster2013identification}, we consider an information-theoretic approach to quantifying similarity between feature vector sequences. One possible approach based on the non-Shannon information measure of Kolmogorov complexity \cite{li2008introduction}, the normalised compression distance (NCD) \cite{li2004similarity}, quantifies similarity between two strings in terms of joint compressibility. The NCD has been applied successfully across a range of problem domains \cite{li2004similarity,kocsor2006application,bardera2010image,wehner2007analyzing}, including music content analysis \cite{li2004melody,cilibrasi2004algorithmic,li2005genre,helen2007similarity,ahonen2009measuring,bello2011measuring}.
For our chosen task of cover song identification, we interpret the NCD as a measure of pairwise predictability. Using our information-theoretic framework, we compare the NCD to alternative predictability measures based on Shannon information. We provide an evaluation of competing information-theoretic approaches and identify issues concerning their implementation. This paper extends our previous work \cite{foster2013identification} as follows: Firstly, we examine a larger set of distance measures and estimate distance measures by predicting discrete-valued sequences. Further, we incorporate the Million Song dataset (MSD) \cite{bertin2011million} into our evaluations. Finally, we investigate combining distance measures using both our considered datasets.

The remainder of this paper is organised as follows: Section~\ref{sec:literaturereview} discusses related work on audio-based cover song identification and methods for determining musical similarity. Section~\ref{sec:background} introduces the pairwise similarity methods evaluated in this work. Section~\ref{sec:method} describes our experimental procedure. Finally, in Sections~\ref{sec:results} and \ref{sec:conclusions} we present results and conclusions.

\section{Related Work}
\label{sec:literaturereview}
\subsection{Musical Similarity}
Methods for characterising similarity between sequences of audio features can be distinguished based on whether the temporal order of features is discarded or retained \cite{casey2008content}. In the former so-called `bag-of-features' approach, a widespread method involves estimating distributions of features obtained from time-frequency representations of musical audio \cite{berenzweig2004large,logan2001music,aucouturier2002music,aucouturier2007bag,mandel2005song,fu2011music,marko2010audio}.
 The bag-of-features approach is unable to model the temporal aspect of music, in which rhythmic, harmonic and melodic objects exhibit sequential structure and in which repetition and variation are of importance \cite{lerdahl1996generative}. Casey and Slaney \cite{casey2006importance} emphasise the role of sequences for music similarity applications, whereas Aucouturier et al.~\cite{aucouturier2007bag} discuss the relative limitations of the bag-of-features approach in a comparison of musical and non-musical audio modelling. Sequential approaches have been utilised in music structure analysis, for identifying repeated and contrasting sequences and their boundaries within a single piece of music \cite{paulus2010state}, in addition to cover song identification.
\subsection{Cover Song Identification}
Owing to the importance of tonal content in determining whether a song is a cover of another, recent cover song identification approaches typically extract representations of the tonal content using chroma features \cite{fujishima1999realtime,bartsch2001catch}.
Chroma features quantify energy distributions across octave-folded bands, using pitch classes in the chromatic scale to map frequency bands to chroma bins.

A variety of cover song identification approaches are based on aligning feature sequences. A widespread approach involves using dynamic programming to determine an optimal set of feature vector insertions, deletions and substitutions, obtained from a similarity matrix. Following Foote's \cite{foote2000arthur} method of applying dynamic time warping (DTW) to a similarity matrix constructed from spectral energy features, G{\'o}mez and Herrera \cite{gomez2006song} propose a DTW approach using chroma features. Serr{\`a} et al.~\cite{serra2008chroma} propose to compute binarised similarity matrices, substituting DTW with an alternative local alignment approach. The cross-recurrence approaches proposed by Serr{\`a} et al.~\cite{serra2009cross} extend the notion of similarity matrices considered in the preceding investigations, in that time-lagged chroma vectors are combined to form higher-dimensional temporal features. In an alternative approach, Serr{\`a} et al.~\cite{serra2012predictability} utilise the previously described method of representing chroma features in combination with non-linear time series prediction techniques, using the cross-prediction error as a measure of similarity.

Using a signal processing approach, Ellis and Poliner \cite{ellis2007identifyingcover} determine component-wise cross-correlation maxima as a measure of similarity between chroma features. Jensen \cite{jensen2008chroma} computes the Euclidean distance between two-dimensional autocorrelations of chroma sequences. More recently, Bertin-Mahieux \cite{bertinlarge} proposes a key-invariant approach based on applying the two-dimensional Fourier transform to chroma sequences.

An alternative approach involves computing similarities between discrete-valued representations of musical content. Tsai et al.~\cite{tsai2005query} apply DTW to discrete-valued sequences, using spectral peak-picking for predominant melody extraction. Bello \cite{bello2007audio} and Lee \cite{lee2006identifying} perform chord estimation with hidden Markov models, using mappings of model states to chords. The resulting sequences are then aligned using DTW. Martin et al.~\cite{martin2012blast} heuristically select chroma bin maxima to determine triads, before locally aligning sequences. We may consider DTW-based approaches, the string-based heuristic evaluated in \cite{martin2012blast} and the cross-correlation approach evaluated in \cite{ellis2007identifyingcover} as alignment techniques, in the sense that they may be used to maximise pairwise correlation between sequences.

With particular regard to this work, a number of approaches are based on applying the NCD to discrete-valued sequences. Using symbolic musical representations directly, Cilibrasi et al.~\cite{cilibrasi2004algorithmic} apply hierarchical clustering to pairwise distances between pieces of music, performing an analysis of clusters with respect to musical genres, musical works and artists. Li and Sleep apply the NCD to genre classification of symbolic musical representations \cite{li2004melody} and musical audio \cite{li2005genre}. 

For audio-based cover song identification, Ahonen \cite{ahonen2009measuring} obtains discrete-valued representations of frame-based chroma features by applying a hidden Markov model (HMM) to perform chord transcription. Predicted chord sequences are then converted to a differential representation, before computing pairwise distances between tracks using the NCD based on different compression algorithms. Ahonen \cite{ahonen2010combining} further proposes to compute multiple discrete-valued representations using additional HMMs and by computing chroma differentials, before averaging separately obtained pairwise distances using the NCD based on prediction by partial matching (PPM) \cite{cleary1984data}. In addition, Ahonen \cite{ahonen2012compression} investigates chroma-derived representations which are compressed using Burrows-Wheeler (BW) compression \cite{burrows1994block}. Bello \cite{bello2011measuring} applies the NCD to recurrence plots computed on individual tracks, as a measure of structural similarity between pieces of music. Finally, Tabus et al.~\cite{tabus2012information} proposes a similar approach to Ahonen based on quantising chroma-derived representations, observing that an alternative compression-based similarity measure outperforms the NCD. Additionally, Silva et al.~\cite{silva2013video} propose a measure of structural similarity based on video compression, observing superior performance using an alternative compression-based measure. Our work extends the above investigations, in that we examine and propose the use of alternative information-theoretic similarity measures to the NCD. Furthermore, we perform an extensive comparison of methods for estimating the NCD and related similarity measures, while proposing approaches which do not require quantising audio features.

A number of recent investigations are concerned with cover song identification using large-scale music collections containing millions of tracks. For such collections, it is typically infeasible to perform computationally expensive pairwise comparisons between a query and every track in the collection. Casey et al.~\cite{casey2008analysis} compute Euclidean distances between windowed chroma sequences. Pairwise similarity is then quantified as the number of distances falling below a threshold. Such an approach may be combined with locality-sensitive hashing  \cite{slaney2008locality} for retrieval with sub-linear time complexity, with respect to a single query. Using a similar approach, Bertin-Mahieux and Ellis \cite{bertin2011large} propose to identify salient `landmark' chroma vectors in individual tracks by applying a thresholding scheme. Identified landmark vectors are then encoded as an integer, thus the collection may be represented as a lookup table. Given a query, the same authors envisage that obtained results are re-ranked using a computationally expensive approach, as proposed by Khadkevich and Omologo \cite{khadkevichlarge}. In this work, we apply such a \textit{filter-and-refine} approach \cite{schnitzer2009filter}, using information-theoretic similarity measures in the refinement stage. 

\subsection{Information-Theoretic Methods} 
Information-theoretic similarity measures  between time series have been proposed in a variety of domains. The idea of jointly compressing two discrete-valued sequences is due to Loewenstern et al.~\cite{loewenstern1995dna} in the context of nucleotide sequence clustering. By parsing sequences using the Lempel-Ziv (LZ) algorithm \cite{ziv1977universal}, Ziv and Merhav \cite{ziv1993measure} propose a method for comparing sequences by compressing one sequence using a model estimated on the other sequence. An alternative approach is considered by Benedetto et al.~\cite{benedetto2002language} for building language trees, where sequences are jointly compressed. Cilibrasi et al.~\cite{cilibrasi2005clustering} motivate their approach of jointly compressing sequences as an approximation of the normalised information distance \cite{li2004similarity}.

\section{Approach}
\label{sec:background}
We denote with $\mat{X} = (\vec{x}_1, \vec{x}_2, \ldots, \vec{x}_N)$, $\mat{Y} = (\vec{y}_1, \vec{y}_2, \ldots, \vec{y}_M)$ two multivariate time series, each representing a sequence of feature vectors extracted from a piece of musical audio. If we assume that both $\mat{X}$, $\mat{Y}$ consist of independent and identically distributed realisations generated respectively by stochastic processes $X$, $Y$, one possible means of quantifying dissimilarity between time series involves the Kullback-Leibler (KL) divergence, defined as
\begin{equation}
\label{eqn:kldivergence}
D_\mathrm{KL}(p_X \| p_Y) = \int p_X(\vec{u}) \log \left(\frac{p_X(\vec{u})}{p_Y(\vec{u})} \right) \, d{\vec{u}}
\end{equation}
where $p_X(\vec{u})$, $p_Y(\vec{u})$ denote the probability density of observation $\vec{u}$ emitted by $X$, $Y$, respectively. Viewed in terms of Shannon information and taking the logarithm to base 2, recall that the KL divergence quantifies the expected number of additional bits required to represent observations emitted by information source $X$, given an optimal code for observations emitted by information source $Y$. The KL divergence has been widely used in conjunction with a `bag-of-features' approach for low-specificity music content analysis tasks \cite{casey2008content}.

To account for temporal structure in musical audio, we may use the NCD as a measure of musical dissimilarity between sequences of quantised feature vectors \cite{li2005genre,ahonen2009measuring,tabus2012information}. Given two strings $x=(x_1, x_2, \ldots, x_N)$, $y=(y_1, y_2, \ldots, y_M)$, the NCD is defined as 
\begin{equation}
\label{eqn:ncd}
\mathrm{NCD}(x,y) = \frac{\max\{C(xy) - C(x), C(yx) - C(y)\}}{\max\{C(x), C(y)\}}
\end{equation}
where $C(\cdot)$ denotes the number of bits required to encode a given string, using a compressor such as the LZ compression algorithm \cite{ziv1977universal}. Similarly, $C(xy)$ denotes the number of bits required to encode the sequential concatenation of strings $x$, $y$. The NCD is an approximation of the normalised information distance (NID) \cite{li2004similarity}, defined as
\begin{equation}
\label{eqn:nid}
\mathrm{NID}(x,y) = \frac{K(x,y) - \min \{K(x),K(y) \}}{\max\{K(x), K(y)\}}
\end{equation}
where the uncomputable function $K(\cdot)$ denotes \textit{algorithmic information content} (AIC), also known as Kolmogorov complexity. The AIC of a given string is the length in bits of the shortest program which outputs the string and then terminates \cite{li2008introduction}. Similarly, $K(x,y)$ denotes the length of the shortest program which outputs $x$, $y$, in addition to a means of distinguishing between both output strings \cite{li2008introduction}. Thus, AIC quantifies the number of bits required to represent specified input strings, under maximally attainable compression. Furthermore, the NID characterises dissimilarity using the transformation under which input strings most closely resemble each other \cite{li2004similarity}.

We are interested in examining the performance of the NCD as an approximation of the NID, where the choice of compressor determines the feature space used to compute similarities \cite{sculley2006compression} in the NCD.
Furthermore, note that the choice of sequential concatenation in $C(xy)$ to approximate $K(x,y)$ represents an additional heuristic \cite{li2004similarity}. In the following sections, we describe our contribution: We first consider in Section~\ref{sec:quantifyingdissimilarityusingshannoninformation} the NID from the perspective of Shannon information, using which we propose a modification to the NCD in Section~\ref{sec:ncda}. We then propose alternative prediction-based measures of similarity in Section~\ref{sec:predictivemodelling}. We detail our approach of applying such measures to continuous-valued sequences in Section~\ref{sec:continuousprediction}.

\subsection{Quantifying Time Series Dissimilarity Using Shannon Information}
\label{sec:quantifyingdissimilarityusingshannoninformation}
We approach the problem of quantifying dissimilarity from the perspective of Shannon information. We assume finite-order, stationary Markov sources $X$, $Y$. We denote with $X_{1:N}$ the sequence of discrete random variables $(X_1, \ldots, X_N)$ emitted by source $X$ at times $1, \ldots, N$. We denote with $H_\mu(X)$, $H_\mu(X,Y)$, $H_\mu(X|Y)$ the entropy rate, joint entropy rate and conditional entropy rate, respectively defined as 
\begin{equation}
\label{eqn:entropyrate}
H_\mu(X) = \lim_{n \to \infty} \frac{1}{n} H(X_1, X_2, \ldots, X_n)
\end{equation}
\begin{equation}
\label{eqn:jointentropyrate}
H_\mu(X,Y) = \lim_{n \to \infty} \frac{1}{n} H((X_1, Y_1), (X_2, Y_2), \ldots, (X_n, Y_n))
\end{equation}
\begin{equation}
\label{eqn:conditionalentropyrate}
H_\mu(X|Y) = H_\mu(X,Y) - H_\mu(Y).
\end{equation}
The entropy rate $H_\mu(X)$ defined in~(\ref{eqn:entropyrate}) quantifies the average amount of uncertainty about $X_n$, while accounting for dependency between $X_n$ for all $n$. Analogously, the joint entropy rate $H_\mu(X,Y)$ defined in~(\ref{eqn:jointentropyrate}) quantifies the average amount of uncertainty about the pair $(X_n,Y_n)$ emitted by sources $X,Y$, while in addition accounting for correlation between the sources. For the conditional entropy rate $H_\mu(X|Y)$ we have
\begin{align}
H_\mu(X|Y) &= \lim_{n \to \infty} \frac{1}{n} H(X_{1:n}, Y_{1:n}) - H(Y_{1:n}) \\
\label{eqn:conditionalentropyidentity}         &= \lim_{n \to \infty} \frac{1}{n} H(X_{1:n}| Y_{1:n}).
\end{align}
From~(\ref{eqn:conditionalentropyidentity}) we may interpret $H_\mu(X|Y)$ as quantifying the average amount of uncertainty about a given emission $X_n$, while taking into account dependency between observations emitted by $X$ and given knowledge of observations emitted by $Y$.

For $N$ observations emitted from source $X$, up to an additive constant the expectation $\mathbf{E}[K(X_{1:N})]$  may be approximated using the entropy \cite{grunwald2004shannon},
\begin{equation}
\mathbf{E}[K(X_{1:N})] \approx  H(X_{1:N}).
\end{equation}
Using~(\ref{eqn:entropyrate}), (\ref{eqn:jointentropyrate}), we assume further approximations
\begin{equation}
\mathbf{E}[K(X_{1:N})] \approx N \,H_\mu(X)
\end{equation}
\begin{equation}
\mathbf{E}[K(X_{1:N}, Y_{1:N})] \approx N \, H_\mu(X,Y)
\end{equation}
where $\mathbf{E}[K(X_{1:N}, Y_{1:N})]$ denotes the expected value of $K(\cdot, \cdot)$ for $N$ observations emitted from sources $X, Y$.
In terms of Shannon information, following \cite{kaltchenko2004algorithms} we use (\ref{eqn:conditionalentropyrate}) and estimate the NID as
\begin{equation}
\label{eqn:shannonnid}
\mathrm{NID}(X,Y) \approx \frac{\max\{H_\mu(X | Y),H_\mu(Y | X)\}}{\max \{H_\mu(X), H_\mu(Y) \}}.
\end{equation}

\subsection{Normalised Compression Distance with Alignment}
\label{sec:ncda}
As given in~(\ref{eqn:shannonnid}), the NID utilises the joint entropy rate $H_\mu(X,Y)$, which accounts for correlation between sources. In contrast, the approach of compressing sequentially concatenated strings to estimate $K(x,y)$ may be inadequate for compressors based on Markov sources, since correlation is not accounted for \cite{kaltchenko2004algorithms}. To address this possible limitation, we propose the normalised compression distance with alignment (NCDA), defined as
\begin{equation}
\label{eqn:ncda}
\mathrm{NCDA}(x,y) = \frac{C(\langle x,y \rangle) - \min \{C(x),C(y) \}}{\max \{C(x),C(y) \}}
\end{equation}
where $\langle a , b \rangle$ performs alignment as a means of maximising correlation between integer-valued strings $a, b$. We generate equal-length strings by padding the shorter of the two strings with the most common value of the longer string. Then, we determine the lag which maximises cross-correlation between strings, before circularly shifting $b$ using the obtained lag value. Finally, we interleave strings. We motivate our choice of cross-correlation by considering that cross-correlation may be computed efficiently, as a series of inner products. Hence, our choice of cross-correlation is pragmatic; an alternative approach might involve minimising NCDA with respect to all lags, or aligning strings using an alternative algorithm.

\subsection{Predictive Modelling}
\label{sec:predictivemodelling}
As previously described, the NCD and NCDA rely on determining the number of bits required to encode strings, using a specified compression algorithm. As an alternative approach, we consider the relation between predictability and compressibility \cite{feder1992universal,feder1994relations} and perform sequence prediction. We illustrate our approach for the case of discrete-valued observations. First, recall that the entropy rate $H_\mu(X)$ is given as
\begin{align}
H_\mu(X) &= \lim_{n \to \infty} - \frac{1}{n} \sum_{x_{1:n} \in \mathcal{A}^n} P_{X}(x_{1:n}) \log P_{X}(x_{1:n}) \label{eqn:jointprobentropyrate}
\end{align}
where $P_{X}(x_{1:n})$ denotes the probability of observing $X_{1:n} = x_{1:n}$, with $x_{1:n} \in \mathcal{A}^n$ according to the alphabet $\mathcal{A}$. We may interpret the quantity $-\log P_{X}(x_{1:n})$ as the number of bits required to represent $\vec{u}_{1:n}$, assuming an optimal code. $H_\mu(X)$ thus quantifies the expected number of bits required to represent a single observation emitted by $X$, while accounting for dependency between observations. Assume that we have an empirical estimate $\hat{P}_{X}$ of the distribution ${P}_{X}$, based on finite observations $x_{1:N}$. Following \cite{begleiter2004prediction}, we estimate $H_\mu(X)$ using \textit{average log-loss} $\ell(\hat{P}_{X},x_{1:N})$, defined as
\begin{align}
\ell(\hat{P}_{X},x_{1:N}) &= - \frac{1}{N} \log \hat{P}_{X}(x_{1:N}) \\
                            &= - \frac{1}{N} \left( \log \hat{P}_{X}(x_1) + \sum_{i=2}^{N} \log \hat{P}_{X}(x_i | x_{1:i-1}) \right) \label{eqn:loglossconditionalprob}
\end{align}
where $\hat{P}_{X}(x_i | x_{1:i-1})$ denotes the estimated probability of observing $x_i$, given preceding context $x_{1:i-1}$. Using~(\ref{eqn:loglossconditionalprob}), we thus compute average log-loss by evaluating the likelihood of observations $x_{1:i-1}$ under the estimated distribution $\hat{P}_{X}$, which we may conceive of as performing a series of predictions based on increasingly long contexts $x_{1:i-1}$. Since $\hat{P}_{X}$ is an estimate of $P_{X}$, the described process is termed \textit{self-prediction} \cite{serra2012predictability}.

We denote with $P_{Y}(x_{1:n})$ the probability of observing $x_{1:n}$ from source $Y$. A measure of disparity between sources $X, Y$ is the cross entropy rate $H_{\mu}^\times(X,Y)$,
\begin{equation}
\label{eqn:crossentropyrate}
H_\mu^\times(X,Y) = \lim_{n \to \infty} - \frac{1}{n} \sum_{x_{1:n} \in \mathcal{A}^n} P_{X}(x_{1:n}) \log P_{Y}(x_{1:n})
\end{equation}
quantifying the expected number of bits required to represent observations emitted by source $X$, given an optimal code for source $Y$. We estimate $H_\mu^\times(X,Y)$ by computing the average log-loss $\ell(\hat{P}_{Y},x_{1:N})$ based on iterated prediction, where $\hat{P}_{Y}$ denotes an estimate of $P_{Y}$ based on observations $y_{1:M}$. Since $\hat{P}_{Y}$, $\hat{P}_{X}$ represent disparate sources, the described process is termed \textit{cross-prediction} \cite{serra2012predictability}. Analogous to the NCD, as a symmetric distance between sources $X,Y$ based on cross entropy, we compute the quantity
\begin{align}
\label{eqn:normalisedcrossentropy}
D^\times(X,Y) &= \frac{H_\mu^\times(X,Y) + H_\mu^\times(Y,X)}{H_\mu(X) + H_\mu(Y)} 
\end{align}
where in~(\ref{eqn:normalisedcrossentropy}) the denominator serves as a normalisation factor, analogous to the denominator in~(\ref{eqn:ncd}) and where we use self-prediction to estimate $H_\mu(X), H_\mu(Y)$.

To obtain a prediction-based estimate of the NID in (\ref{eqn:shannonnid}), we may estimate $H_\mu(X)$, $H_\mu(Y)$ again using self-prediction. Furthermore, we estimate the conditional entropy rate $H_\mu(X|Y)$ using the distribution $\hat{P}_{X|Y}$, referring to the estimated distribution of observations emitted by $X$, given knowledge of observations $y_{1:M}$ emitted by $Y$. Analogous to self-prediction and cross-prediction, we define the quantity $\ell(\hat{P}_{X|Y}, x_{1:N}, y_{1:M})$,
\begin{equation}
\label{eqn:conditionalselfprediction}
\begin{split}
\ell(\hat{P}_{X|Y},x_{1:N}, y_{1:M}) = \\ - \frac{1}{N} \left( \log \hat{P}_{X|Y}(x_1 | y_{1:M}) + \sum_{i=2}^{N} \log \hat{P}_{X|Y}(x_i | x_{1:i-1}, y_{1:M}) \right).
\end{split}
\end{equation}
We refer to the process used to compute (\ref{eqn:conditionalselfprediction}) as \textit{conditional self-prediction}.

\begin{figure*}[tb]
\opt{preprint}{\begin{minipage}[b]{.50\linewidth}}
\opt{journal}{\begin{minipage}[b]{.33\linewidth}}
  \centering
  \centerline{\includegraphics[width=5.7cm]{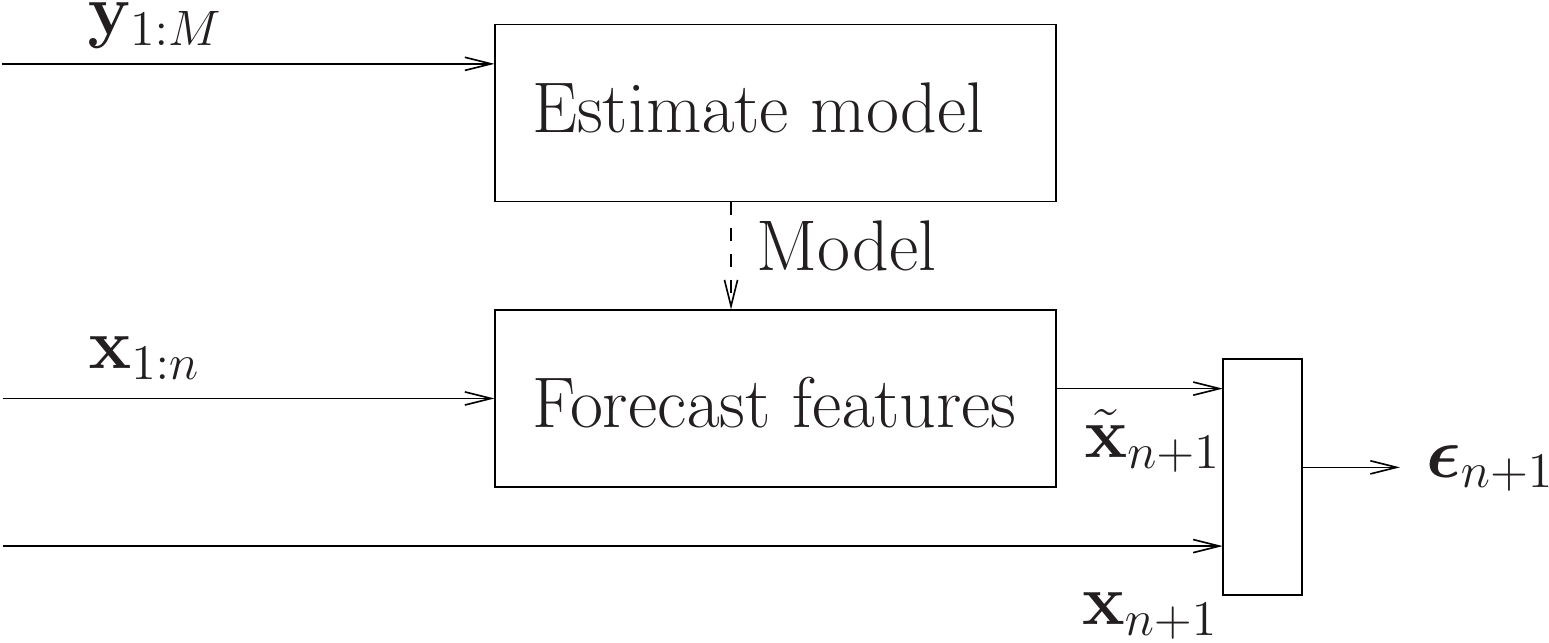}}
  \centerline{(a) Cross-prediction}\medskip
  \label{fig:crossprediction}
\end{minipage}
\opt{preprint}{\begin{minipage}[b]{.50\linewidth}}
\opt{journal}{\begin{minipage}[b]{.33\linewidth}}
  \centering
  \centerline{\includegraphics[width=5.7cm]{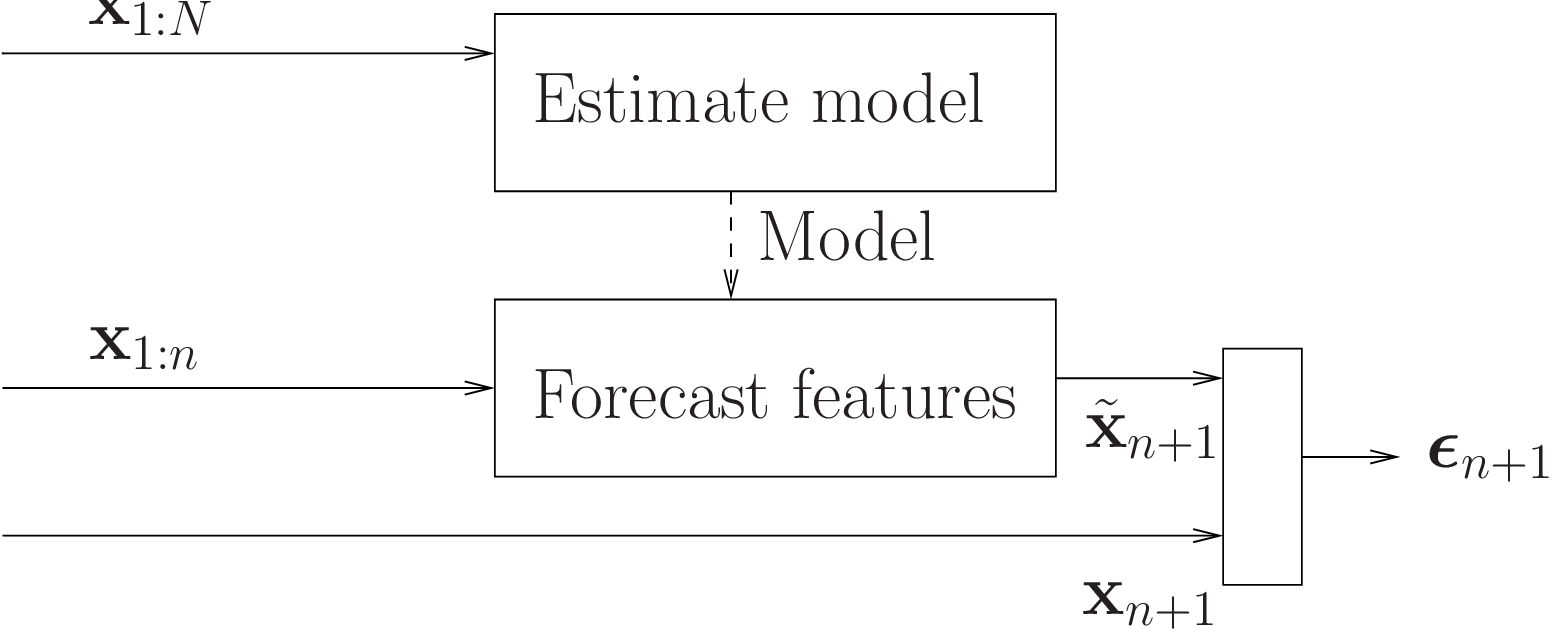}}
  \centerline{(b) Self-prediction}\medskip
  \label{fig:selfprediction}
\end{minipage}
\opt{preprint}{\begin{minipage}[b]{.99\linewidth}}
\opt{journal}{\begin{minipage}[b]{.33\linewidth}}
  \centering
  \centerline{\includegraphics[width=5.7cm]{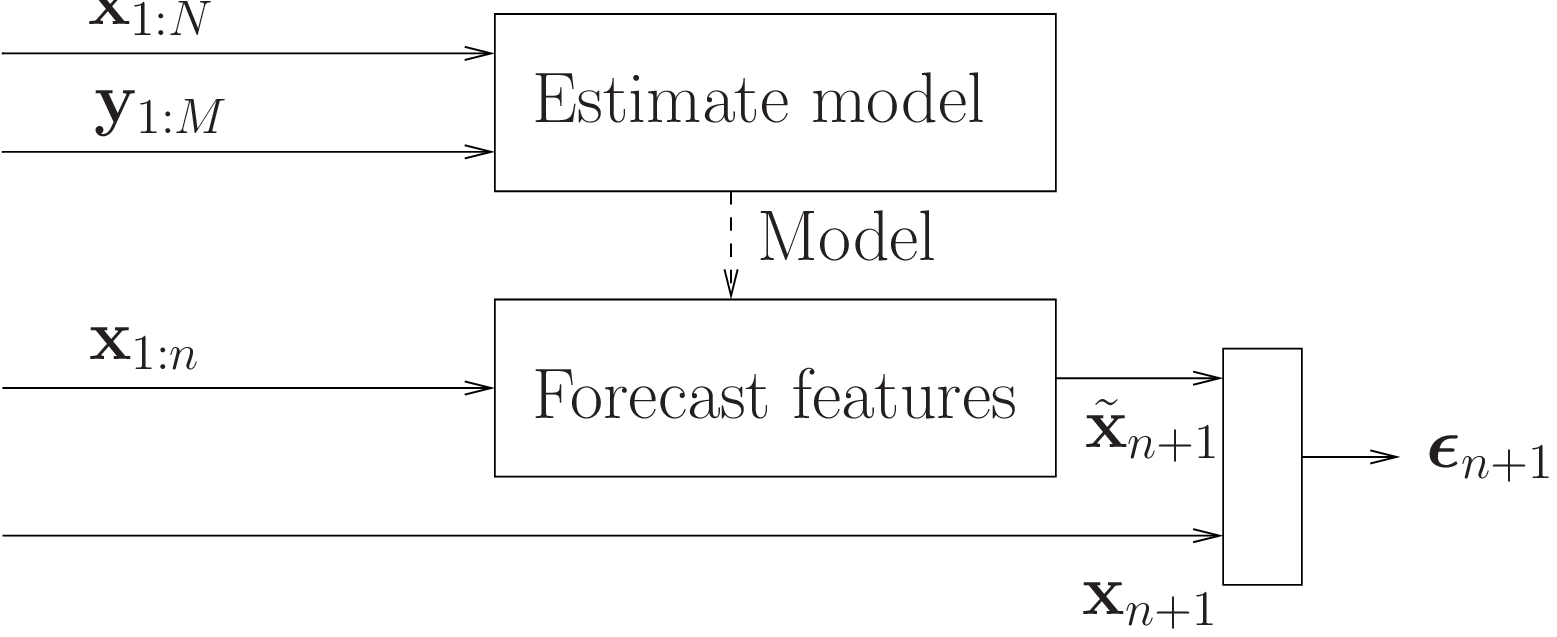}}
  \centerline{(c) Conditional self-prediction}\medskip
  \label{fig:conditionalselfprediction}
\end{minipage}
\caption{Evaluated prediction strategies. Sequences $\vec{x}_{1:N}, \vec{y}_{1:M}$ serve as model inputs, observation context $\vec{x}_{1:n}$ forms basis of prediction $\tilde{\vec{x}}_{n+1}$. Quantity $\greekvec{\epsilon}_{n+1}$ denotes prediction error.}
\label{fig:predictiontypes}
\end{figure*}

\subsection{Continuous-Valued Approach}
\label{sec:continuousprediction}
The quantities described in Section \ref{sec:predictivemodelling} may be computed  using quantised feature vectors \cite{li2005genre,ahonen2009measuring,marko2010audio,tabus2012information}. As an alternative, we propose an approach requiring no prior quantisation.
As used in \cite{serra2012predictability}, in our approach we utilise non-linear time series prediction. In contrast to \cite{serra2012predictability}, we are concerned with evaluating distance measures which we compute as statistics of prediction errors. Therefore, we use a comparatively straightforward nearest-neighbours approach. Given the sequence of feature vectors $\mathbf{C}$, consider first the process of \textit{time-delay embedding} \cite{takens1981detecting}, which yields the vector sequence $\vec{S}^\vec{C}$, whose elements $\vec{s}_r^{\mathbf{C}}$ are defined as
\begin{align}
\label{eqn:statespaceembedding}
\vec{s}^{\mathbf{C}}_{r} &= \mathrm{vec} \, (\vec{c}_{r}, \vec{c}_{(r-1) \tau}, \ldots, \vec{c}_{(r-d+1) \tau}).
\end{align}
According to~(\ref{eqn:statespaceembedding}), each element $\vec{s}_r^{\mathbf{C}}$ aggregates feature vector $\vec{c}_r$ along with its preceding temporal context $(\vec{c}_{(r-1) \tau}, \ldots, \vec{c}_{(r-d+1) \tau})$. The amount of temporal context is controlled by parameters $d$, $\tau$, respectively referred to as \textit{embedding dimension} and \textit{time delay}. Operator $\mathrm{vec}$ denotes vectorisation.

Our method of predicting features is based on determining nearest neighbours in time-delay embedded space. We first illustrate our method for the case of cross-prediction, depicted schematically in Fig.~\ref{fig:predictiontypes}~(a). Given sequence $\vec{y}_{1:M}$, we denote with $\tilde{\vec{x}}_{t+h}$ the estimated successor of sequence $\vec{x}_{1:t+h-1}$,
\begin{equation}
\tilde{\vec{x}}_{t+h} = \vec{y}_{q(t)+h}
\end{equation}
where $h$ denotes the \textit{predictive horizon} (how far into the future we predict), and where we define $q(t)$ as
\begin{equation}
\label{eqn:crossprediction}
q(t) =\argmax_{k \in [d\,..\,M-h]} \mathrm{corr}(\vec{s}_k^\mathbf{Y}, \vec{s}_t^\mathbf{X})
\end{equation}
with $\mathrm{corr}(\vec{s}_k^\mathbf{Y}, \vec{s}_t^\mathbf{X})$ denoting the sample Pearson correlation coefficient between vectors $\vec{s}_k^\mathbf{Y}, \vec{s}_t^\mathbf{X}$. We motivate use of correlation coefficients as an alternative to the Euclidean distance, following \cite{gomez2006tonal}.

Depicted schematically in Fig.~\ref{fig:predictiontypes}~(b), to perform self-prediction we set $\mathbf{Y} = \mathbf{X}$. Since features may be slowly-varying, when forming prediction $\tilde{\vec{x}}_{t+h}$ we disregard observations in the immediate past of time step $t$. Thus we define
\begin{equation}
\tilde{\vec{x}}_{t+h} = \vec{x}_{q'(t)+h}
\end{equation}
with $q'(t)$ defined as
\begin{equation}
\label{eqn:nearestneighbourselfprediction}
q'(t) = \argmax_{k \in [d\,..\,N-h], \, |k-t| > R} \mathrm{corr}(\vec{s}_k^\mathbf{X}, \vec{s}_t^\mathbf{X})
\end{equation}
and where $R$ denotes the radius below which observations are disregarded.

Finally, to perform conditional self-prediction, we use both time-delay embedded spaces $\vec{s}^\mathbf{Y}$, $\vec{s}^{\mathbf{X}}$. Given predictions $\vec{y}_{q(t)+h}$, $\vec{x}_{q'(t)+h}$, respectively obtained using cross-prediction and self-prediction, we compute the linear combination
\begin{equation}
\label{eqn:nearestneighbourconditionalselfprediction}
\tilde{\vec{x}}_{t+h} = \vec{y}_{q(t)+h} \, \alpha + \vec{x}_{q'(t)+h} \, (1-\alpha).
\end{equation}
Similar to the approach given in \cite{foster2011causal}, in (\ref{eqn:nearestneighbourconditionalselfprediction}) for weighting coefficient $\alpha$ we use
\begin{equation}
\alpha = \frac{\mathrm{MSE}_\mathrm{self}}{\mathrm{MSE}_\mathrm{self} + \mathrm{MSE}_\mathrm{cross}}
\end{equation}
where $\mathrm{MSE}_\mathrm{cross}$, $\mathrm{MSE}_\mathrm{self}$ respectively denote cross-prediction and self-prediction mean squared errors. Fig.~\ref{fig:predictiontypes}~(c) depicts conditional self-prediction schematically.

Given the sequence of predictions $\tilde{\vec{x}}_{1:N}$, we denote with $\greekvec{\epsilon}_n$ the rescaled prediction error, whose $i$th component $\epsilon_{i,n}$ is given by
\begin{equation}
\epsilon_{i,n} = \frac{\tilde{{x}}_{i,n}-{x}_{i,n}}{s_i}
\end{equation}
where $s_i$ denotes the sample variance of the $i$th component $(\vec{x}_{1:N})_i$ in $\vec{x}_{1:N}$. We contrast our approach with the component-wise normalised mean squared error (NMSE) based on cross-prediction used in \cite{serra2012predictability}, which may be applied as an alternative measure of dissimilarity between time series. Our approach is based on assuming that the prediction error may be represented using a normally distributed random variable $Z$ with samples $\greekvec{\epsilon}_{1:N}$. Using the samples, we estimate the prediction error entropy $H(Z)$ parametrically. In the case of self-prediction, we assume the approximation $H(Z) \approx H_\mu(X)$; analogously in the case of cross-prediction and conditional self-prediction, we assume respective approximations $H(Z) \approx H_\mu^\times(X,Y)$, $H(Z) \approx H_\mu(X|Y)$. Assuming normality, we estimate $H(Z)$ using the equation
\begin{equation}
H(Z) = \frac{1}{2} \log(2 \pi e)^k |\mathbf{\Sigma}|
\end{equation}
where $\mathbf{\Sigma}$ denotes the sample covariance of $Z$. In our continuous-valued approach, using the prediction methods depicted in Fig.~\ref{fig:predictiontypes}, we thus estimate information-based measures as statistics of the prediction error sequence. We then substitute the obtained quantities in (\ref{eqn:shannonnid}) and (\ref{eqn:normalisedcrossentropy}) to obtain continuous-valued, prediction-based analogues of the NID and distance $D^\times$. The continuous-valued, prediction-based approach contrasts with our discrete-valued, prediction-based methods previously described in Section~\ref{sec:predictivemodelling} and our discrete-valued, compression-based method described in Section~\ref{sec:ncda}.

\section{Experimental Method}
\label{sec:method}

We first evaluate our proposed methods using a set of $300$ audio recordings of Jazz standards\,\footnote{\url{http://www.eecs.qmul.ac.uk/~peterf/jazzdataset.html}}.
We assume that two tracks are a cover pair if they possess identical title strings. Thus, we assume a symmetric relation when determining cover identities. The equivalence class of tracks deemed to be covers of one another is a \textit{cover set}. The Jazz data set comprises 97 cover sets, with average cover set size 3.06 tracks.

Furthermore, we perform a large-scale evaluation based on the MSD \cite{bertin2011million}. This dataset includes meta-data and pre-computed audio features for a collection of $10^6$ Western popular music recordings. We use a pre-defined evaluation set of 5\,236 query tracks partitioned into 1\,726 cover sets\,\footnote{\url{http://labrosa.ee.columbia.edu/millionsong/secondhand}}, with average cover set size 3.03 tracks. Following \cite{bertinlarge}, for each query track, we seek to identify the remaining cover set members contained in the entire $10^6$ track collection.

\subsection{Feature Extraction}
For the Jazz dataset, as a representation of musical harmonic content, we extract 12-component beat-synchronous chroma features from audio using the method and implementation described in \cite{ellis2007identifyingcover}.
Assuming an equal-tempered scale, the method accounts for deviations in standard pitch from 440Hz, by shifting the mapping of FFT bins to pitches in the range of $\pm 0.5$ semitones. Following chroma extraction, beat-synchronisation is achieved using the method described in \cite{ellis2006beat}. First, onset detection is performed by differencing a log-magnitude Mel-frequency spectrogram across time and applying half-wave rectification, before summing across frequency bands. After high-pass filtering the onset signal, a tempo estimate is formed by applying a window function to the autocorrelated onset signal and determining autocorrelation maxima. Varying the centre of the window function allows tempo estimation to incorporate a bias towards a preferred beat rate (PBR).
The tempo estimate and onset signal are then used to obtain an optimal set of beat onsets, by using dynamic programming. Chroma features are averaged over beat intervals, before applying square-root compression and normalising chroma features with respect to the Euclidean norm. Based on our previous work \cite{foster2013identification}, we evaluate using a PBR of 240 beats per minute (bpm).

The MSD includes 12-component chroma features alongside predicted note and beat onsets \cite{echonestfeatures}, which we use in our evaluations. In contrast to the beat-synchronous features obtained for the Jazz dataset, MSD chroma features are initially aligned to predicted onsets. Motivated by our choice of PBR for the Jazz dataset, we resample predicted beat onsets to a rate of 240bpm. We then average chroma features over resampled beat intervals. Finally, we normalise features as described for the Jazz dataset.

\subsection{Key Invariance}
To account for musical key variation within cover sets, we transpose chroma sequences using the optimal transposition index (OTI) method \cite{serra2008chroma}. Given two chroma vector sequences $\mathbf{X}$, $\mathbf{Y}$, we form summary vectors $\vec{h}_\mathbf{X}$, $\vec{h}_\mathbf{Y}$ by averaging over entire sequences. The OTI corresponds to the number of circular shift operations applied to $\vec{h}_\mathbf{Y}$ which maximises the inner product between $\vec{h}_\mathbf{X}$ and $\vec{h}_\mathbf{Y}$,
\begin{equation}
\mathrm{OTI}(\vec{h}_\mathbf{X},\vec{h}_\mathbf{Y}) = \argmax_i \, \vec{h}_\mathbf{X} \cdot \mathrm{circshift}(\vec{h}_\mathbf{Y},i)
\end{equation}
where $\mathrm{circshift}(\vec{h}_\mathbf{Y},i)$ denotes applying $i$ circular shift operations to $\vec{h}_\mathbf{Y}$. We subsequently shift chroma vectors  $\mathbf{Y}$ by $\mathrm{OTI}(\vec{h}_\mathbf{X},\vec{h}_\mathbf{Y})$ positions, prior to pairwise comparison. 

\subsection{Quantisation}
\label{sec:discretisation}
For discrete-valued similarity measures, we quantise chroma features using the $K$-means algorithm. We cluster chroma features aggregated across all tracks, where we consider codebook sizes in the range $[2\,..\,48]$. To increase stability, we execute the $K$-means algorithm $20$ times. We then select the clustering which minimises the mean squared error between data points and assigned clusters. The described quantisation method performs similarly to an alternative based on pairwise sequence quantisation; for a detailed discussion we refer to our previous work \cite{foster2013identification}.

\subsection{Distance Measures}
\label{sec:distancemeasures}
We summarise the distance measures evaluated in this work in Table~\ref{tab:distancemeasures}, where for each distance measure, we list our estimation methods. 

We utilise the following algorithms to compute distance measures by compressing strings: Prediction by partial matching (PPM) \cite{cleary1984data}, Burrows-Wheeler (BW) compression \cite{burrows1994block} and Lempel-Ziv (LZ) compression \cite{ziv1977universal}, implemented respectively as PPMD\,\footnote{\url{http://compression.ru/ds/}}, BZIP2\,\footnote{\url{http://bzip2.org}} and ZLIB\,\footnote{\url{http://zlib.org}}.
In all cases, we set parameters to favour compression rates over computation time. To obtain strings, following quantisation we map integer codewords to alphanumeric characters.

We use the described compression algorithms to determine the length in bits of compressed strings and compute NCD, NCDA distances. In a complementary discrete-valued approach, we use string prediction instead of compression. Using average log-loss, we compute NCDA using the formula
\begin{equation}
\frac{\ell(\hat{P}_{\langle X,Y \rangle}, \langle x,y \rangle) - \min \{\ell(\hat{P}_X, x ), \ell(\hat{P}_Y, y) \}}{\max\{\ell(\hat{P}_X,x),\ell(\hat{P}_Y,y) \}}
\end{equation}
where $\ell(\hat{P}_{\langle X,Y \rangle},  \langle x,y \rangle)$ is the average log-loss obtained from performing self-prediction on the aligned sequence $\langle x,y \rangle$. We compute a prediction-based variant of NCD analogously by predicting sequentially concatenated strings without performing any alignment. In addition, we use cross-prediction to estimate distance measure $D^\times$, as defined in~(\ref{eqn:normalisedcrossentropy}). We perform string prediction using Begleiter's \cite{begleiter2004prediction} implementations of PPMC and LZ78 algorithms.

Note that the KL divergence given in~(\ref{eqn:kldivergence}) is non-symmetric. In our evaluations, we observed that computing a symmetric distance improved performance; based on KL divergence, we compute the Jensen-Shannon divergence (JSD) $D_{\mathrm{JS}}(p_X \| p_Y)$, defined as
\begin{equation}
\label{eqn:jsd}
D_\mathrm{JS}(p_X \| p_Y) = D_\mathrm{KL}(p_X \| p_A) + D_\mathrm{KL}(p_Y \| p_A)
\end{equation}
where $p_A$ denotes the mean of $p_X, p_Y$,
\begin{equation}
p_A = \frac{1}{2} \left(p_X + p_Y\right).
\end{equation}
As a baseline method, we compute the JSD between symbol histograms normalised to sum to one.

We evaluate continuous-valued prediction using time-delay embedding parameters $h \in \{ 1,4 \}$, $d \in \{1,2,4\}$, $\tau \in \{1,2,4,6\}$, setting the exclusion radius in (\ref{eqn:nearestneighbourselfprediction}) to $R=8$ based on preliminary analysis using separate training data. We compute distance measure $D^\times$ using cross-prediction to estimate the numerator in (\ref{eqn:normalisedcrossentropy}). In a complementary approach, we estimate the NID using conditional self-prediction to estimate the numerator in (\ref{eqn:shannonnid}). For $D^\times$ and NID, we use self-prediction to estimate the denominator in (\ref{eqn:normalisedcrossentropy}), (\ref{eqn:shannonnid}), respectively.

Finally, to compensate for cover song candidates consistently deemed similar to query tracks, we normalise pairwise distances using the method described in \cite{ravuri2010cover}. We apply distance normalisation as a post-processing step, before computing performance statistics.

\begin{table*}
\centering
\begin{tabular}{c l l l}
\hline
\hline
Distance \opt{journal}{measure} & Definition & \multicolumn{2}{l}{Estimation method} \\
\hline
$\mathrm{NCD}$      & Eqn.~\ref{eqn:ncd} & {\small String compression (LZ, BW, PPM)}& \opt{journal}{{\small Discrete prediction (LZ, PPM)}} \\ %
\opt{preprint}{ & & {\small Discrete prediction (LZ, PPM)} & \\}
$\mathrm{NCDA}$     & Eqn.~\ref{eqn:ncda} & {\small String compression (LZ, BW, PPM)} & \opt{journal}{{\small Discrete prediction (LZ, PPM)}} \\
\opt{preprint}{ & & {\small Discrete prediction (LZ, PPM)} & \\}
$D^\times$          & Eqn.~\ref{eqn:normalisedcrossentropy}   & {\small Discrete prediction} & \opt{journal}{{\small Continuous prediction}} \\
\opt{preprint}{ & & {\small Continuous prediction} & \\}
$D_\mathrm{JS}$     & Eqn.~\ref{eqn:jsd} & \multicolumn{2}{l}{\small Normalised symbol histograms}  \\
$\mathrm{NID}$      & Eqn.~\ref{eqn:shannonnid} & \multicolumn{2}{l}{\small Continuous prediction} \\
\hline

\end{tabular}
\caption{Summary of evaluated distance measures.}
\label{tab:distancemeasures}
\end{table*}

\subsection{Large-scale Cover Song Identification}
For music content analysis involving large datasets, algorithm scalability is an important issue. The approaches in this work by themselves require a linear scan through the dataset for a given query, which may be infeasible for large datasets. We use a scalable approach for our evaluations involving the MSD. Following \cite{khadkevichlarge} and similar to the method proposed in \cite{osmalskyj2013efficient}, we incorporate our methods into a two-stage retrieval process. By using a metric distance to determine similarity in the first retrieval stage, we allow for the potential use of indexing or hashing schemes, as proposed in \cite{casey2008analysis,schnitzer2009filter}. We then apply non-metric pairwise comparisons in the second retrieval stage.

In the first stage, we quantise as described in Section~\ref{sec:discretisation} and represent each track with a normalised codeword histogram. Given a query track, we then rank each of the $10^6$ candidate tracks using the L1 distance. To account for key variation, for each candidate track we minimise L1 distance across chroma rotations. We then determine the top $L=1000$ candidate tracks, which we re-rank in the second stage using our proposed methods. After both retrieval stages, we normalise pairwise distances as described in Section~\ref{sec:distancemeasures}. We report performance based on the final ranking of all $10^6$ candidate tracks, across query tracks.

\subsection{Performance Statistics}
\label{sec:performancestatistics}
As used in \cite{bello2011measuring}, we quantify cover song identification accuracy using mean average precision (MAP), based on ranking tracks according to distance with respect to queries.
The MAP is obtained by averaging query-wise scores, where we may interpret each score as the average of precision values at the ranks of relevant tracks, where relevant tracks in our case are covers of the query track.
Following \cite{bello2011measuring}, we use the Friedman test \cite{friedman1937use} to compare accuracies among distance measures. The Friedman test is based on ranking across queries each distance measure according to average precision. We combine the Friedman test with Tukey's range test \cite{tukey1973problem} to adjust for Type I errors when performing multiple comparisons.

As a subsidiary performance measure, for each query we compute the precision at rank $r$, with $r \in \{5, 10, 20\}$. We subsequently average across queries to obtain mean precision at rank $r$.

\subsection{Combining Distance Measures}
\label{sec:combiningdistances}
To determine if combining distance measures improves cover song identification accuracy, we obtain pairwise distances as described in Section \ref{sec:distancemeasures}. We denote with $d^k_{i,j}$ the pairwise distance between the $i$th query track and the $j$th result candidate, obtained using the $k$th distance measure in our evaluation. We transform $d^k_{i,j}$ by computing the inverse rank $d\,'^k_{i,j}$,
\begin{equation}
  d\,'^k_{i,j} = 1 - {\mathrm{rank}( d^k_{i,j})}^{-1}
\end{equation}
where $\mathrm{rank}(d^k_{i,j})$ denotes the rank of $d^k_{i,j}$ among all distances obtained with respect to query track $i$, given the $k$th distance measure. We apply this transformation to protect against outliers, while ensuring that distance decreases rapidly for track pairs deemed highly similar, for decreasing distance. Note that since our distance transformation preserves monotonicity and MAP itself is based on ranked distances, performance of unmixed distance measures is uninfluenced by this transformation. Finally, we combine distances $d\,'^k_{i,j}$, $d\,'^m_{i,j}$ by computing a weighted average of distances pooled using $\max$ and $\min$ operators,
\begin{equation}
\label{eqn:combiningoperators}
\max \{ d\,'^k_{i,j}, d\,'^m_{i,j} \} \, \beta + \min \{ d\,'^k_{i,j},  d\,'^m_{i,j} \} \, (1 - \beta)
\end{equation}
where we vary $\beta$ in the range $[0,1]$. We motivate our approach on the basis that we may interpret inverse ranks as estimated probabilities of cover identities, furthermore the operators $\max$ and $\min$ have been proposed as a means of combining probability estimates for classification \cite{kittler1998combining}. In forming a linear combination, we evaluate the utility of $\max$ pooling versus $\min$ pooling. An alternative approach based on straightforward averaging did not yield any performance gain.

\subsection{Baseline Approaches}
In addition to the JSD and cross-prediction NMSE baselines, we include an evaluation of the method and implementation described in \cite{ellis2007identifyingcover} based on cross-correlation. As a random baseline, we sample pairwise distances from a normal distribution.

\section{Results}
\label{sec:results}
\subsection{Discrete-Valued Approaches Based on Compression}

\begin{figure*}
\begin{minipage}[b]{\opt{journal}{.25\linewidth}\opt{preprint}{.50\linewidth}}
  \centerline{\includegraphics[width=4.0cm]{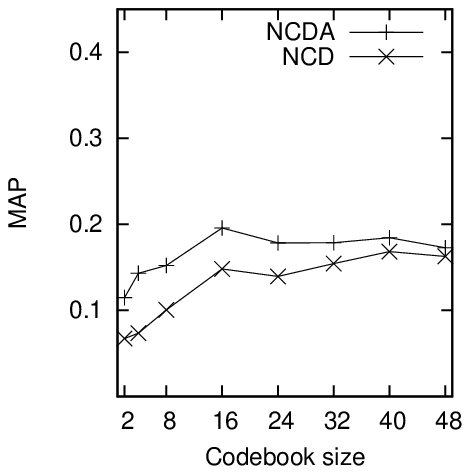}}
  \centerline{(a) LZ (Jazz)}\medskip
\end{minipage}%
\begin{minipage}[b]{\opt{journal}{.25\linewidth}\opt{preprint}{.50\linewidth}}
  \centerline{\includegraphics[width=4.0cm]{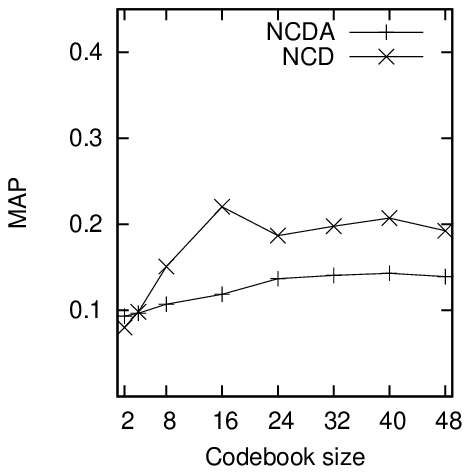}}
  \centerline{(b) BW (Jazz)}\medskip
\end{minipage}
\begin{minipage}[b]{\opt{journal}{.25\linewidth}\opt{preprint}{.50\linewidth}}
  \centering
  \centerline{\includegraphics[width=4.0cm]{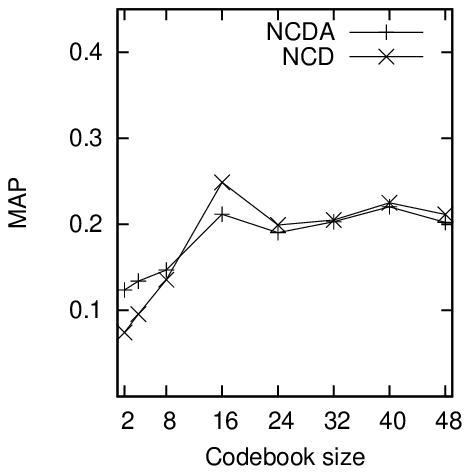}}
  \centerline{(c) PPM (Jazz)}\medskip
\end{minipage}%
\begin{minipage}[b]{\opt{journal}{.25\linewidth}\opt{preprint}{.50\linewidth}}
  \centering
  \centerline{\includegraphics[width=4.0cm]{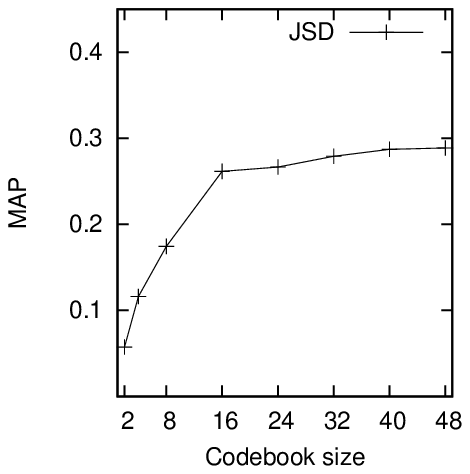}}
  \centerline{(d) JSD (Jazz)}\medskip
\end{minipage}
\begin{minipage}[b]{\opt{journal}{.25\linewidth}\opt{preprint}{.50\linewidth}}
  \centerline{\includegraphics[width=4.0cm]{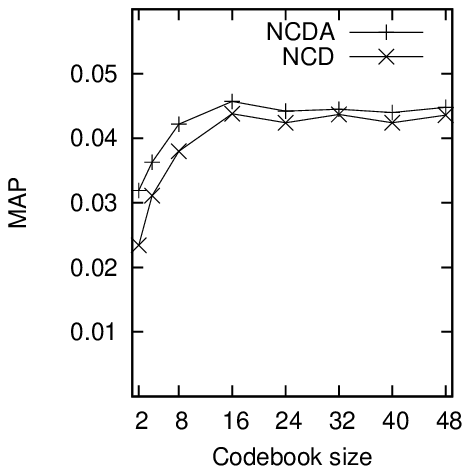}}
  \centerline{(e) LZ (MSD)}\medskip
\end{minipage}%
\begin{minipage}[b]{\opt{journal}{.25\linewidth}\opt{preprint}{.50\linewidth}}
  \centerline{\includegraphics[width=4.0cm]{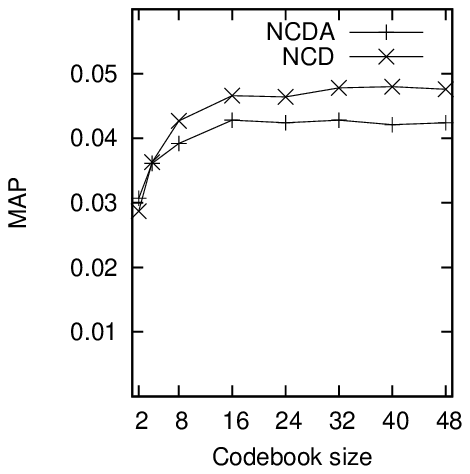}}
  \centerline{(f) BW (MSD)}\medskip
\end{minipage}
\begin{minipage}[b]{\opt{journal}{.25\linewidth}\opt{preprint}{.50\linewidth}}
  \centering
  \centerline{\includegraphics[width=4.0cm]{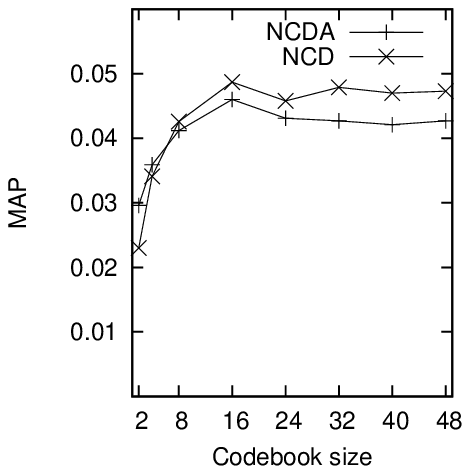}}
  \centerline{(g) PPM (MSD)}\medskip
\end{minipage}%
\begin{minipage}[b]{\opt{journal}{.25\linewidth}\opt{preprint}{.50\linewidth}}
  \centering
  \centerline{\includegraphics[width=4.0cm]{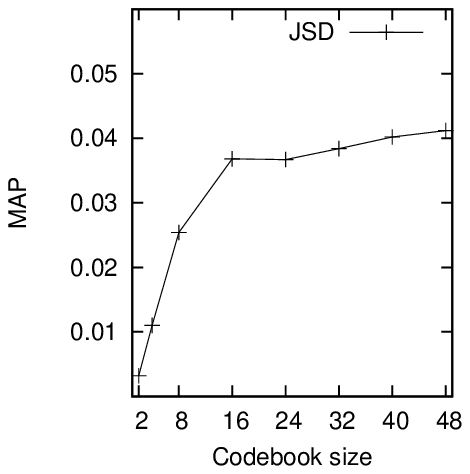}}
  \centerline{(h) JSD (MSD)}\medskip
\end{minipage}
%
%
%
\caption{Effect of codebook size and distance measure on mean average precision (MAP). Results displayed for Lempel-Ziv (LZ), Burrows-Wheeler (BW) and prediction by partial matching (PPM) algorithms in subfigures (a)--(c), (e)--(g), for Jazz and MSD datasets respectively. Subfigures (d), (h) display results for Jensen-Shannon divergence baseline (JSD), for Jazz and MSD datasets respectively.}
\label{fig:stringcompressionresults}
\end{figure*}

In Fig.~\ref{fig:stringcompressionresults}~(a)--(c), we examine the performance of discrete-valued NCD and NCDA distance measures, combined with LZ, BW and PPM algorithms and based on the Jazz dataset. For the LZ algorithm, NCDA yields a relative performance gain of $38.6$\%, averaged across codebook sizes. In contrast, for PPM, with the exception of small codebook sizes in the range $[2\,..\,8]$, NCDA yields no consistent improvement over NCD, however averaged across codebook sizes we obtain a mean relative performance gain of $11.0$\%. Finally, the effect of using NCDA is reversed for BW compression, where performance decreases by an average of $21.8$\%. 

Examining results for the MSD in Fig.~\ref{fig:stringcompressionresults}~(e)--(g), we observe similar qualitative results for LZ and BW algorithms. For the LZ algorithm, NCDA yields an average relative performance gain of $10.1$\%, whereas for BW compression we observe an average relative performance loss of $6.5$\%. In contrast to the Jazz dataset, for PPM we observe an average relative performance loss of $1.5$\%.

For both datasets, NCDA appears to be most advantageous combined with LZ compression, whereas BW yields the least advantageous result. Note that BW compression is block-based in contrast to LZ and PPM compressors, both of which are sequential.
We attribute this observation to performance differences among compressors, since the assumptions made in Section~\ref{sec:ncda} rely on assuming Markov sources. Noting differences in relative performance gains between datasets, following \cite{khadkevichlarge} we further conjecture that chroma feature representation influences the performance of the evaluated distance measures.

We examine the performance of JSD between normalised symbol histograms, as displayed in Fig.~\ref{fig:stringcompressionresults}~(d), (h). Surprisingly, for the Jazz dataset and for $K>8$, JSD outperforms compression-based methods, with maximum MAP score $0.289$ obtained for $K=48$. This result is contrary to our expectation that NCD approaches should outperform the bag-of-features approach, by accounting for temporal structure in time series. In contrast, for the MSD and for optimal $K$, both NCD and NCDA outperform JSD across all evaluated compression algorithms. We attribute this disparity to differences in dataset size, where for the Jazz dataset the problem size may be sufficiently small to amortise advantages of using NCD, NCDA compared to JSD.

\subsection{Discrete-Valued Approaches Based on Prediction}
\begin{figure}[tb]
\begin{minipage}[b]{.48\linewidth}
  \centering
  \centerline{\includegraphics[width=4.0cm]{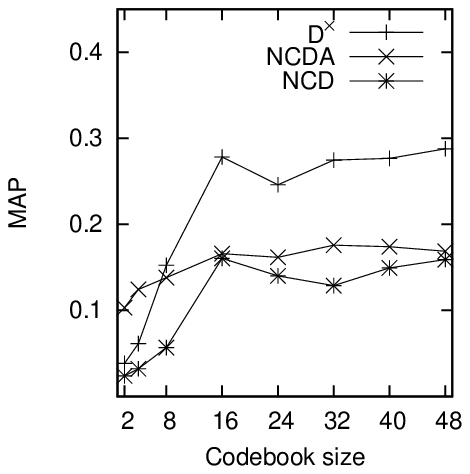}}
  \centerline{(a) LZ (Jazz)}\medskip
\end{minipage}%
\begin{minipage}[b]{.48\linewidth}
  \centering
  \centerline{\includegraphics[width=4.0cm]{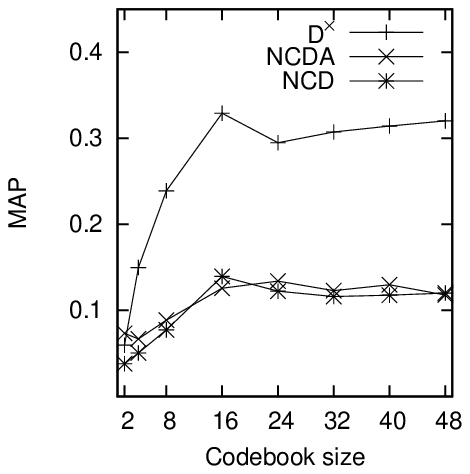}}
  \centerline{(b) PPM (Jazz)}\medskip
\end{minipage}
\begin{minipage}[b]{.48\linewidth}
  \centering
  \centerline{\includegraphics[width=4.0cm]{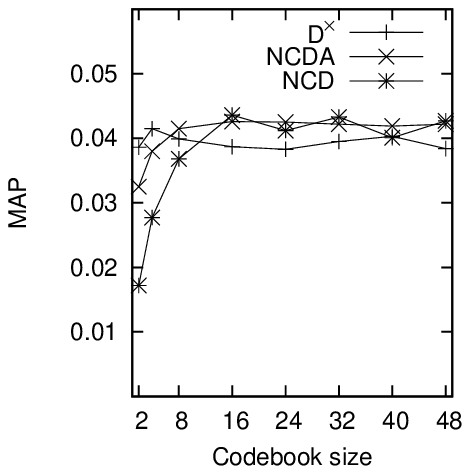}}
  \centerline{(c) LZ (MSD)}\medskip
\end{minipage}%
\begin{minipage}[b]{.48\linewidth}
  \centering
  \centerline{\includegraphics[width=4.0cm]{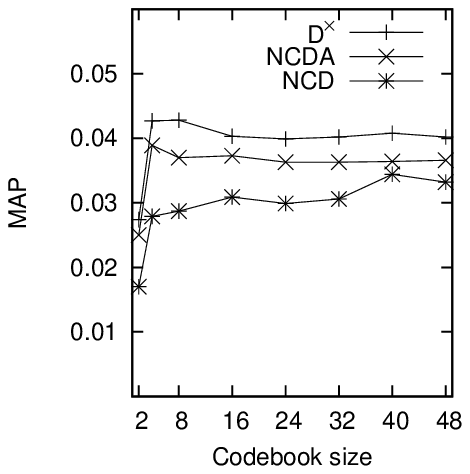}}
  \centerline{(d) PPM (MSD)}\medskip
\end{minipage}
%
\caption{Effect of codebook size and distance measure on mean average precision (MAP). Results obtained using string prediction approach, displayed for Lempel-Ziv (LZ) (subfigures (a), (c)) and prediction by partial match (PPM) (subfigures (b), (d)), for Jazz and MSD datasets respectively.}
\label{fig:stringpredictionresults}
\end{figure}

In Fig.~\ref{fig:stringpredictionresults}, we consider the performance of distance measures based on string prediction. For the Jazz dataset, comparing log-loss estimates of NCD and NCDA using the LZ algorithm, averaged across codebook sizes NCDA outperforms NCD; we obtain a mean relative performance gain of $105.1\%$ (Fig.~\ref{fig:stringpredictionresults} (a)). For the PPM algorithm, although NCD maximises performance (MAP $0.140$), we obtain a mean relative performance gain of $19.3$\% using NCDA over NCD (Fig.~\ref{fig:stringpredictionresults} (b)). Importantly, for both LZ and PPM the cross-prediction distance $D^\times$ consistently outperforms NCD and NCDA; for $K=16$ and combined with PPM compression, we obtain MAP $0.329$. For the MSD and using LZ compression, in contrast to the Jazz dataset we observe a mean relative performance loss of $1.8$\% when comparing $D^\times$ with NCDA. For both LZ and PPM, NCDA compared to NCD yields mean relative performance gains of $17.6$\% and $24.0$\%, respectively.

\begin{table*}
  \begin{minipage}[b]{\opt{journal}{.34}\opt{preprint}{.99}\linewidth}
    \scriptsize
    \centering
    \begin{tabular}{ l|l|cccc }
               & $d$ & $\tau=1$ & $\tau=2$ & $\tau=4$ & $\tau=6$ \\ \hline
    \multirow{3}{*}{h=1}
     & 1 & 0.282 &  0.282 &  0.282 &  0.282 \\
     & 2 & 0.308  &  0.311  &  0.293  &  0.312  \\
     & 4 & 0.327  &  0.332  &  0.318  &  0.318  \\
     \hline
    \multirow{3}{*}{h=4}
     & 1 & 0.243  &  0.243  &  0.243  &  0.243  \\
     & 2 & 0.262  &  0.273  &  0.291  &  0.284  \\
     & 4 & 0.307  &  0.313  & \textbf{0.346} &  0.321  \\ 
    \end{tabular}
	\centerline{ \small (a) NID estimate; conditional self-prediction (Jazz)}\medskip
  \end{minipage}
  \begin{minipage}[b]{\opt{journal}{.34}\opt{preprint}{.99}\linewidth}
    \scriptsize
    \centering
    \begin{tabular}{ l|l|cccc }
               & $d$ & $\tau=1$ & $\tau=2$ & $\tau=4$ & $\tau=6$ \\ \hline
    \multirow{3}{*}{h=1}
     & 1 & 0.347  &  0.347  &  0.347  &  0.347   \\
     & 2 & 0.412  &  0.403  &  0.390  &  0.403    \\
     & 4 & \textbf{0.454} &  0.446  &  0.432  &  0.423    \\
     \hline
    \multirow{3}{*}{h=4}
     & 1 & 0.293  &  0.293  &  0.293  &  0.293   \\
     & 2 & 0.352  &  0.364  &  0.377  &  0.365   \\
     & 4 & 0.408  &  0.428  &  0.432  &  0.435   \\ 
    \end{tabular}
    \centerline{ \small (b) $D^\times$ estimate; cross-prediction (Jazz)}\medskip
  \end{minipage}
  \begin{minipage}[b]{\opt{journal}{.34}\opt{preprint}{.99}\linewidth}
    \scriptsize
    \centering
    \begin{tabular}{ l|l|cccc }
               & $d$ & $\tau=1$ & $\tau=2$ & $\tau=4$ & $\tau=6$ \\ \hline
    \multirow{3}{*}{h=1}
     & 1 & 0.344  &  0.344  &  0.344  &  0.344   \\
     & 2 & 0.402  &  0.396  &  0.385  &  0.389   \\
     & 4 & 0.448  &  0.452  &  0.428  &  0.433   \\
     \hline
    \multirow{3}{*}{h=4}
     & 1 & 0.321  &  0.321  &  0.321  &  0.321   \\
     & 2 & 0.362  &  0.375  &  0.390  &  0.379   \\
     & 4 & 0.417  &  0.450  &  0.446  &  \textbf{0.459} \\ 
    \end{tabular}
    \centerline{ \small (c) NMSE; cross-prediction (Jazz)}\medskip
  \end{minipage}
  \begin{minipage}[b]{\opt{journal}{.34}\opt{preprint}{.99}\linewidth}
    \scriptsize
    \centering
    \begin{tabular}{ l|l|cccc }
               & $d$ & $\tau=1$ & $\tau=2$ & $\tau=4$ & $\tau=6$ \\ \hline
    \multirow{3}{*}{h=1}
     & 1 & 0.0191 &  0.0191  & 0.0191 &  0.0191 \\
     & 2 & 0.0230 &  0.0222  & 0.0239 &  0.0250 \\
     & 4 & 0.0238 &  0.0275  & \textbf{0.0303} &  0.0295 \\
     \hline
    \multirow{3}{*}{h=4}
     & 1 & 0.0200 &  0.0200 &  0.0200 &  0.0200 \\
     & 2 & 0.0208 &  0.0239 &  0.0236 &  0.0260 \\
     & 4 & 0.0228 &  0.0276 &  0.0303 &  0.0301 \\ 
    \end{tabular}
	\centerline{ \small (d) NID estimate; conditional self-prediction (MSD)}\medskip
  \end{minipage}
  \begin{minipage}[b]{\opt{journal}{.34}\opt{preprint}{.99}\linewidth}
    \scriptsize
    \centering
    \begin{tabular}{ l|l|cccc }
               & $d$ & $\tau=1$ & $\tau=2$ & $\tau=4$ & $\tau=6$ \\ \hline
    \multirow{3}{*}{h=1}
     & 1 & 0.0451 &  0.0451 &  0.0451 &  0.0451 \\
     & 2 & 0.0476 &  0.0477 &  0.0479 &  0.0475 \\
     & 4 & 0.0489 &  0.0494 &  0.0494 &  0.0489 \\
     \hline
    \multirow{3}{*}{h=4}
     & 1 & 0.0465 &  0.0465 &  0.0465 &  0.0465 \\
     & 2 & 0.0470 &  0.0480 &  0.0484 &  0.0487 \\
     & 4 & 0.0478 &  0.0488 &  \textbf{0.0498} &  0.0491 \\ 
    \end{tabular}
    \centerline{ \small (e) $D^\times$ estimate; cross-prediction (MSD)}\medskip
  \end{minipage}
  \begin{minipage}[b]{\opt{journal}{.34}\opt{preprint}{.99}\linewidth}
    \scriptsize
    \centering
    \begin{tabular}{ l|l|cccc }
               & $d$ & $\tau=1$ & $\tau=2$ & $\tau=4$ & $\tau=6$ \\ \hline
    \multirow{3}{*}{h=1}
     & 1 & 0.0341 &  0.0341 &  0.0341 &  0.0341 \\
     & 2 & 0.0404 &  0.0420 &  0.0431 &  0.0437 \\
     & 4 & 0.0447 &  0.0474 &  0.0478 &  0.0465 \\
     \hline
    \multirow{3}{*}{h=4}
     & 1 & 0.0431 &  0.0431 &  0.0431 &  0.0431 \\
     & 2 & 0.0450 &  0.0457 &  0.0467 &  0.0471 \\
     & 4 & 0.0466 &  0.0494 &  \textbf{0.0499} & 0.0494 \\ 
    \end{tabular}
    \centerline{ \small (f) NMSE; cross-prediction (MSD)}\medskip
  \end{minipage}
  \caption{Mean average precision scores for distances based on continuous prediction. In each subfigure, parameters $h$, $\tau$, $d$ denote predictive horizon, time delay and embedding dimension, respectively. Results displayed in subfigures (a)--(c), (d)--(f) for Jazz and MSD datasets, respectively.}
  \label{tab:continuousresults}
\end{table*}

\subsection{Continuous-Valued Approaches}
Table~\ref{tab:continuousresults} displays the performance of continuous-valued prediction approaches. Note that for $d=1$, parameter $\tau$ may be set to an arbitrary integer following (\ref{eqn:statespaceembedding}). We consider results obtained for the Jazz dataset (Table~\ref{tab:continuousresults} (a)--(c)). Using conditional self-prediction to estimate the NID, maximised across parameters $h, d, \tau$ we obtain MAP $0.346$. In comparison, cross-prediction distance $D^\times$ yields MAP $0.454$. As a baseline, we determine the cross-prediction NMSE, where maximising across parameters we obtain MAP $0.459$. Table~\ref{tab:continuousresults} (a)--(c) displays performance against evaluated parameter combinations. Examining results for the MSD in Table~\ref{tab:continuousresults} (d)--(f), we obtain qualitatively similar results with maximum MAP values $0.0303$, $0.0498$ and $0.0499$ for NID, $D^\times$ and NMSE, respectively. For both datasets, we observe that increasing the value of $d$ consistently improves performance. In contrast, we observe no such effect for parameters $\tau, h$.

\subsection{Summary of Results and Comparison to State of the Art}
Fig.~\ref{fig:significancetesting} (a), (b) displays the result of significance testing as described in Section~\ref{sec:performancestatistics}, where we assume 95\% confidence intervals and where we maximise across evaluated parameter spaces. Table~\ref{tab:singlepredictors} displays a corresponding summary of MAP scores. As baselines we include Ellis and Poliner's cross-correlation approach \cite{ellis2007identifyingcover}, in addition to randomly sampled pairwise distances. For the MSD, when used without any further refinement method, our filtering stage based on normalised codeword histograms yields MAP 0.0056.

For both Jazz dataset and MSD, we observe that continuous-valued approaches based on cross-prediction consistently outperform discrete-valued approaches. Moreover, with the exception of NCD combined with PPM-based string compression and for the MSD, using continuous-valued cross-prediction significantly outperforms discrete-valued approaches. For approaches based on string compression, we note that using NCDA with BW compression significantly decreases performance with respect to NCD. Similarly, using NCDA decreases MAP scores for PPM. Although we do not observe a significant performance gain using NCDA over NCD for LZ compression, performance improves consistently across datasets. For the Jazz dataset, we observe that the JSD baseline significantly outperforms the majority of string-compression approaches. In contrast, for the MSD the majority of string-compression approaches significantly outperform the JSD baseline. Whereas PPM with distance $D^\times$ consistently outperforms all discrete-valued approaches for the Jazz dataset, PPM with compression-based NCD consistently outperforms all discrete-valued approaches for the MSD and significantly outperforms the JSD baseline. 

In a comparison of continuous-valued approaches, we observe that cross-prediction using either distance $D^\times$ or NMSE competes with cross-correlation for the Jazz dataset. In contrast, the same cross-prediction approaches significantly outperform cross-correlation for the MSD.

Examining continuous-valued approaches further, for both Jazz dataset and MSD, we observe a significant disadvantage in using our conditional self-prediction based estimate of NID, over cross-prediction based distances $D^\times$ and NMSE. The relatively poor performance of NID for the MSD suggests a limitation of our prediction approach when used with MSD chroma features. However, considering results for both datasets suggests that cross-prediction yields more favourable results than conditional self-prediction generally.

To facilitate further comparison, we consider the approaches proposed by Bertin-Mahieux and Ellis \cite{bertinlarge}, Khadkevich and Omologo \cite{khadkevichlarge}, who report MAP scores of 0.0295, 0.0371, respectively. Based on such a comparison, we obtain state-of-the-art results. Note that the stated approaches do not report any distance normalisation procedure as described in Section \ref{sec:distancemeasures}; we found that normalisation improves our results: For the Jazz dataset and using unnormalised distances, we obtain MAP scores $0.425$, $0.314$, $0.332$ for NMSE, $D^\times$, NID, respectively. For the MSD and using unnormalised distances, we obtain MAP scores $0.0340$, $0.0174$, $0.0216$, for NMSE, $D^\times$, NID, respectively.

\subsection{Combining Distances}
\begin{figure}
  \begin{minipage}[b]{0.99\linewidth}
    \centering
    \centerline{\includegraphics[width=8.3cm]{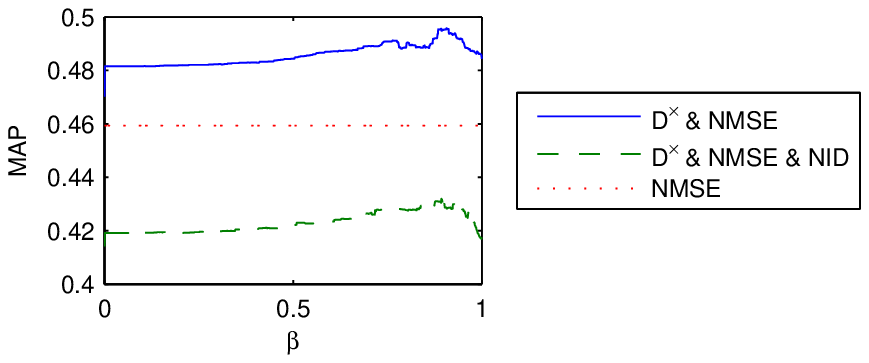}}
    \centerline{(a) Jazz}
  \end{minipage}
  \begin{minipage}[b]{0.99\linewidth}
    \centering
    \centerline{\includegraphics[width=8.3cm]{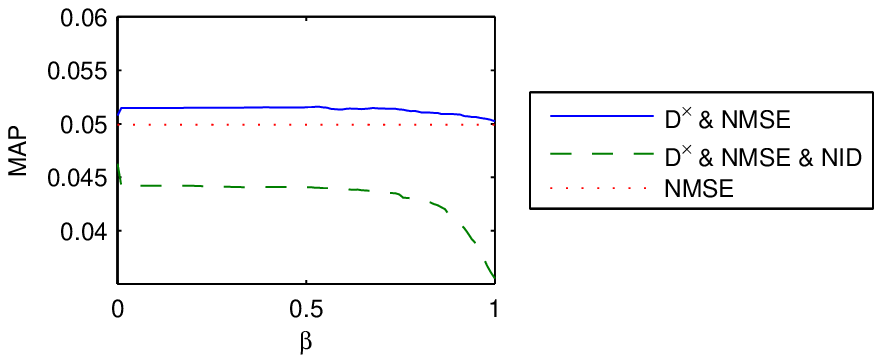}}
    \centerline{(b) MSD}
  \end{minipage}
  \caption{Mean average precision for combinations of distances, in response to parameter $\beta$. Results displayed for Jazz dataset and MSD in subfigures (a) and (b), respectively.}
  \label{fig:mapversusbeta}
\end{figure}

Finally, using the method described in Section~\ref{sec:combiningdistances}, we combine distances obtained using continuous-valued prediction. Fig.~\ref{fig:mapversusbeta} displays MAP scores against mixing parameter $\beta$, for Jazz dataset and MSD. We consider the combinations $D^\times \&\,\text{NMSE}$, $D^\times \&\,\text{NMSE}\,\&\,\text{NID}$, the latter combination which we evaluate with respect to optimal $\beta$ for the former combination.

Compared to using the baseline NMSE alone, across all $\beta$ and for both datasets we observe that combining NMSE with $D^\times$ improves performance: For the Jazz dataset, we observe maximal MAP score 0.496, corresponding to a gain of $8.1$\%. For the MSD, we observe maximal MAP score $0.0516$, corresponding to a gain of $3.4\%$. We observe no performance gain by further combining NID estimates with NMSE and $D^\times$, obtaining maximal MAP scores 0.432 and 0.0463 respectively for Jazz dataset and MSD. Additional evaluations revealed no performance gain using unnormalised distances.

\begin{figure}
  \begin{minipage}[b]{0.60\linewidth}
    \centerline{\includegraphics[width=5.3cm,angle=-90]{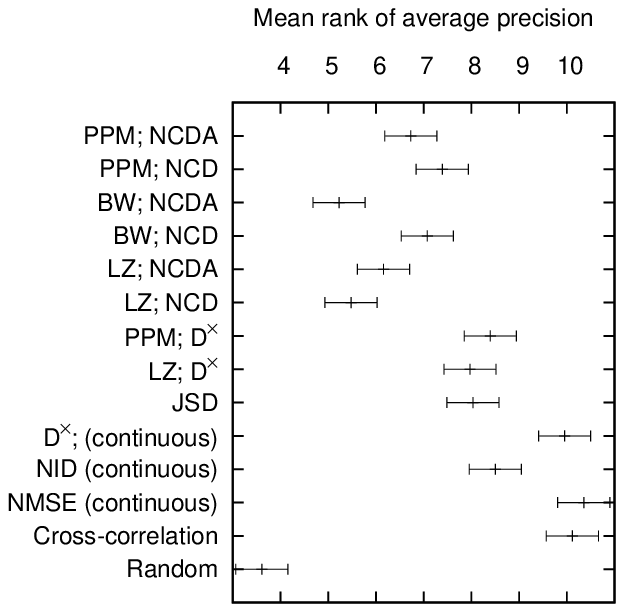}}
	\vspace{0.2cm}
    \centerline{ \hspace{1.2cm} (a) Jazz}
  \end{minipage}
  \begin{minipage}[b]{0.38\linewidth}
    \centerline{\includegraphics[width=5.3cm,angle=-90]{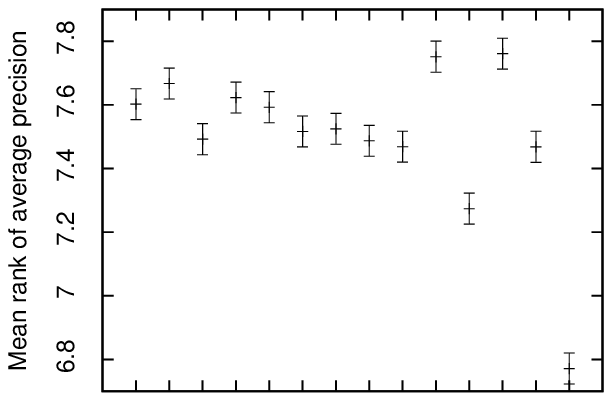}}
	\vspace{0.2cm}
    \centerline{(b) MSD}
  \end{minipage}
  \begin{minipage}[b]{0.60\linewidth}
    \vspace{0.2cm}
    \centerline{\includegraphics[width=3.15cm, angle=-90]{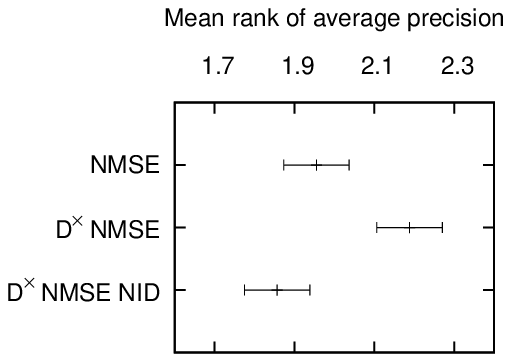}}
	\vspace{0.2cm}
    \centerline{ \hspace{1.2cm} (c) Jazz}
  \end{minipage}
  \begin{minipage}[b]{0.38\linewidth}
    \vspace{0.2cm}
    \centerline{\includegraphics[width=3.15cm, angle=-90]{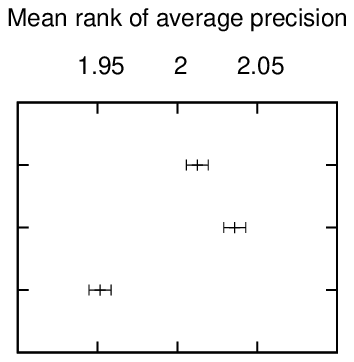}}
	\vspace{0.2cm}
    \centerline{(d) MSD}
  \end{minipage}
  \caption{Mean ranks of average precision scores obtained using Friedman test. Error bars indicate 95\% confidence intervals obtained using Tukey's range test \cite{tukey1973problem}. Higher mean ranks indicate higher performance. Results displayed for Jazz and MSD datasets in subfigures (a) and (b), respectively, with results for combined distances displayed in subfigures (c) and (d).}
  \label{fig:significancetesting}
\end{figure}

Table~\ref{tab:singlepredictors} summarises MAP scores; in Fig.~\ref{fig:significancetesting} (c),~(d) we test for differences in performance among combinations of distances based on continuous-valued prediction. Compared to using the baseline NMSE alone, combining NMSE with $D^\times$ significantly improves performance for both the Jazz dataset and MSD. In addition, Table~\ref{tab:precisionatr} reports performance in terms of mean precision at ranks $r$. Matching previous observations, for Jazz dataset and MSD, the combination of NMSE and $D^\times$ consistently outperforms remaining combinations. At rank $r=5$, relative to the NMSE baseline, we obtain a performance gain of $10.0\%$ for the Jazz dataset and $6.7\%$ for the MSD.

\begin{table*}[tc]
\footnotesize
\centering
\begin{tabular}{c|cc|cc}
Dataset & \multicolumn{2}{c}{Jazz} & \multicolumn{2}{c}{MSD} \\
\hline
Method  & NCDA & NCD & NCDA & NCD \\
\hline
PPM   & 0.220 $\pm$ 0.021 & 0.249 $\pm$ 0.021  & 0.0460 $\pm$ 0.0024 & 0.0487 $\pm$ 0.0025 \\ 
BW  & 0.143 $\pm$ 0.016 & 0.220 $\pm$ 0.019 & 0.0428 $\pm$ 0.0023 & 0.0480 $\pm$  0.0024\\  
LZ   & 0.196 $\pm$ 0.019  & 0.168 $\pm$ 0.017 & 0.0457 $\pm$ 0.0024 & 0.0438 $\pm$ 0.0023 \\ 
\hline
PPM; $D^{\times}$ &\multicolumn{2}{c|}{0.329 $\pm$ 0.022 } & \multicolumn{2}{c}{0.0428 $\pm$ 0.0022} \\ 
LZ; $D^{\times}$ &\multicolumn{2}{c|}{0.288 $\pm$ 0.021} & \multicolumn{2}{c}{0.0415 $\pm$ 0.0022} \\ 
JSD &\multicolumn{2}{c|}{0.289 $\pm$ 0.022} & \multicolumn{2}{c}{0.0412 $\pm$ 0.0023} \\
$D^{\times}$ (continuous) &\multicolumn{2}{c|}{0.454 $\pm$ 0.024} & \multicolumn{2}{c}{0.0498 $\pm$ 0.0025} \\ 
NID (continuous) &\multicolumn{2}{c|}{0.346 $\pm$ 0.023} & \multicolumn{2}{c}{0.0303 $\pm$ 0.0020} \\
NMSE (continuous) &\multicolumn{2}{c|}{0.459 $\pm$ 0.023} & \multicolumn{2}{c}{0.0499 $\pm$ 0.0025} \\
Ellis and Poliner \cite{ellis2007identifyingcover}& \multicolumn{2}{c|}{0.465 $\pm$ 0.024} & \multicolumn{2}{c}{0.0404 $\pm$ 0.0023} \\ 
Random &\multicolumn{2}{c|}{0.026 $\pm$ 0.004} & \multicolumn{2}{c}{0.0006 $\pm$ 0.0001} \\
\hline
$D^{\times}$ \& NMSE (cont.) & \multicolumn{2}{c|}{0.496} & \multicolumn{2}{c}{0.0516 $\pm$ 0.0025} \\
$D^{\times}$ \& NID \& NMSE (cont.) & \multicolumn{2}{c|}{0.432} & \multicolumn{2}{c}{0.0463 $\pm$ 0.0024} \\
\end{tabular}
\caption{Summary of mean average precision scores. First three rows denote compression based approaches. Intervals are standard errors. `Random' denotes sampling pairwise distances from a normal distribution.}
\label{tab:singlepredictors}
\end{table*}

\begin{table*}[tc]
\footnotesize
\centering
\begin{tabular}{c|ccc|ccc}
Dataset & \multicolumn{3}{c}{Jazz} & \multicolumn{3}{c}{MSD} \\
\hline
$r$ & 5 & 10 & 20 & 5 & 10 & 20 \\
\hline
$D^{\times}$ & 0.185 &  0.113 &  0.065  & 0.0276 &  0.0146 &  0.0077 \\ 
NID & 0.133  &  0.075  &  0.045  & 0.0147 &  0.0082 &  0.0044   \\
NMSE & 0.193  &  0.116  &  0.067  & 0.0270 &  0.0141 &  0.0075 \\
\hline
$D^{\times}$ \& NMSE & 0.213  & 0.123  &  0.070  & 0.0288 &  0.0150  & 0.0079\\
$D^{\times}$ \& NID \& NMSE & 0.168  &  0.101  &  0.063  & 0.0265 &  0.0146  & 0.0076 \\
\end{tabular}
\caption{Mean precision at rank $r$, for approaches based on continuous-valued prediction.}
\label{tab:precisionatr}
\end{table*}

\section{Conclusions}
\label{sec:conclusions}
We have evaluated measures of pairwise predictability between time series for cover song identification. We consider alternative distance measures to the NCD: We propose NCDA, which incorporates a method for obtaining joint representations of time series, in addition to methods based on cross-prediction. Secondly, we attend to the issue of representing time series: We propose continuous-valued prediction as a means of determining pairwise similarity, where we estimate compressibility as a statistic of the prediction error. We contrast methods requiring feature quantisation, against methods directly applicable to continuous-valued features.

Firstly, the proposed continuous-valued approach outperforms discrete-valued approaches and competes with evaluated continuous baseline approaches. Secondly, we draw attention to using cross-prediction as an alternative approach to the NCD, where we observe superior results in both discrete and continuous cases for Jazz cover song identification, and for the continuous case for cover song identification using the Million Song Dataset. Thirdly, using NCDA, we are able to mitigate differences in performance between evaluated discrete compression algorithms. We view the previous three points as evidence that using information-based measures of similarity, a continuous-valued representation may be preferable to discrete-valued chroma representations, owing to the challenge of obtaining discrete-valued representations. Further, NCD may yield suboptimal performance compared to alternative distance measures. 

We argue that due to the ubiquity of time series similarity problems, our results are relevant to application domains extending beyond the scope of this work. Finally, in the context of cover song identification, we have demonstrated state-of-the-art performance using a large-scale dataset. We have shown that our distances based on continuous-valued prediction may be combined to improve performance relative to the baseline.

For future work, we aim to evaluate alternative time series models to those presently considered. To this end, further investigations might involve causal state space reconstruction \cite{shalizi2004blind} or recurrent neural networks such as the long short term memory architecture \cite{hochreiter1997long}. For future work, we aim to evaluate ensemble techniques for combining distances in greater detail.


\opt{journal}{
  \ifCLASSOPTIONcaptionsoff
    \newpage
  \fi
}

\bibliographystyle{IEEEtran}
\bibliography{main}

\begin{thebibliography}{10}
\providecommand{\url}[1]{#1}
\csname url@samestyle\endcsname
\providecommand{\newblock}{\relax}
\providecommand{\bibinfo}[2]{#2}
\providecommand{\BIBentrySTDinterwordspacing}{\spaceskip=0pt\relax}
\providecommand{\BIBentryALTinterwordstretchfactor}{4}
\providecommand{\BIBentryALTinterwordspacing}{\spaceskip=\fontdimen2\font plus
\BIBentryALTinterwordstretchfactor\fontdimen3\font minus
  \fontdimen4\font\relax}
\providecommand{\BIBforeignlanguage}[2]{{%
\expandafter\ifx\csname l@#1\endcsname\relax
\typeout{** WARNING: IEEEtran.bst: No hyphenation pattern has been}%
\typeout{** loaded for the language `#1'. Using the pattern for}%
\typeout{** the default language instead.}%
\else
\language=\csname l@#1\endcsname
\fi
#2}}
\providecommand{\BIBdecl}{\relax}
\BIBdecl

\bibitem{casey2008content}
M.~A. Casey, R.~Veltkamp, M.~Goto, M.~Leman, C.~Rhodes, and M.~Slaney,
  ``Content-based music information retrieval: current directions and future
  challenges,'' \emph{Proc.~IEEE}, vol.~96, no.~4, pp. 668--696, 2008.

\bibitem{cano2005review}
P.~Cano, E.~Batlle, T.~Kalker, and J.~Haitsma, ``A review of audio
  fingerprinting,'' \emph{Journal VLSI Signal Process.}, vol.~41, no.~3, pp.
  271--284, 2005.

\bibitem{serra2011identification}
J.~Serr{\`a}, ``Identification of versions of the same musical composition by
  processing audio descriptions,'' Ph.D. dissertation, Universitat Pompeu
  Fabra, 2011.

\bibitem{fujihara2010modeling}
H.~Fujihara, M.~Goto, T.~Kitahara, and H.~G. Okuno, ``A modeling of singing
  voice robust to accompaniment sounds and its application to singer
  identification and vocal-timbre-similarity-based music information
  retrieval,'' \emph{IEEE Trans.~Audio, Speech, Language Process.}, vol.~18,
  no.~3, pp. 638--648, 2010.

\bibitem{mandel2005song}
M.~I. Mandel and D.~P.~W. Ellis, ``Song-level features and support vector
  machines for music classification,'' in \emph{Proc.~6th Intern.~Conf.~Music
  Information Retrieval (ISMIR)}, 2005, pp. 594--599.

\bibitem{scaringella2006automatic}
N.~Scaringella, G.~Zoia, and D.~J. Mlynek, ``Automatic genre classification of
  music content: a survey,'' \emph{IEEE Signal Process.~Magazine}, vol.~23,
  no.~2, pp. 133--141, 2006.

\bibitem{serra2008chroma}
J.~Serr{\`a}, E.~G{\'o}mez, P.~Herrera, and X.~Serra, ``Chroma binary
  similarity and local alignment applied to cover song identification,''
  \emph{IEEE Trans.~Audio, Speech, Language Process.}, vol.~16, no.~6, pp.
  1138--1151, 2008.

\bibitem{meyer1956music}
L.~Meyer, \emph{Music and Emotion}.\hskip 1em plus 0.5em minus 0.4em\relax
  University of Chicago Press, 1956.

\bibitem{shannon1949mathematical}
C.~E. Shannon, ``{A mathematical theory of communication},'' \emph{Bell System
  Technical Journal}, vol.~27, pp. 379--423 623--656, 1948.

\bibitem{huron2006sweet}
D.~Huron, \emph{Sweet Anticipation: Music and the Psychology of
  Expectation}.\hskip 1em plus 0.5em minus 0.4em\relax MIT press, 2006.

\bibitem{pearce2012auditory}
M.~T. Pearce and G.~A. Wiggins, ``Auditory expectation: The information
  dynamics of music perception and cognition,'' \emph{Topics in Cognitive
  Science}, vol.~4, no.~4, 2012.

\bibitem{abdallah2009information}
S.~Abdallah and M.~D. Plumbley, ``Information dynamics: patterns of expectation
  and surprise in the perception of music,'' \emph{Connection Science},
  vol.~21, no. 2-3, pp. 89--117, 2009.

\bibitem{foster2013identification}
P.~Foster, S.~Dixon, and A.~Klapuri, ``Identification of cover songs using
  information theoretic measures of similarity,'' in \emph{Proc.~IEEE
  Intern.~Conf.~Acoustics, Speech, and Signal Process.~(ICASSP)}, 2013.

\bibitem{li2008introduction}
M.~Li and P.~Vit{\'a}nyi, \emph{An Introduction to Kolmogorov Complexity and
  its Applications}.\hskip 1em plus 0.5em minus 0.4em\relax Springer, 2008.

\bibitem{li2004similarity}
M.~Li, X.~Chen, X.~Li, B.~Ma, and P.~M.~B. Vit{\'a}nyi, ``The similarity
  metric,'' \emph{IEEE Trans.~Inf.~Theory}, vol.~50, no.~12, pp. 3250--3264,
  2004.

\bibitem{kocsor2006application}
A.~Kocsor, A.~Kert{\'e}sz-Farkas, L.~Kaj{\'a}n, and S.~Pongor, ``Application of
  compression-based distance measures to protein sequence classification: a
  methodological study,'' \emph{Bioinformatics}, vol.~22, no.~4, pp. 407--412,
  2006.

\bibitem{bardera2010image}
A.~Bardera, M.~Feixas, I.~Boada, and M.~Sbert, ``Image registration by
  compression,'' \emph{Information Sciences}, vol. 180, no.~7, pp. 1121--1133,
  2010.

\bibitem{wehner2007analyzing}
S.~Wehner, ``Analyzing worms and network traffic using compression,''
  \emph{Journal of Computer Security}, vol.~15, no.~3, pp. 303--320, 2007.

\bibitem{li2004melody}
M.~Li and R.~Sleep, ``Melody classification using a similarity metric based on
  {K}olmogorov complexity,'' \emph{Sound and Music Computing}, pp. 126--129,
  2004.

\bibitem{cilibrasi2004algorithmic}
R.~Cilibrasi, P.~M.~B. Vit{\'a}nyi, and R.~Wolf, ``Algorithmic clustering of
  music based on string compression,'' \emph{Computer Music Journal}, vol.~28,
  no.~4, pp. 49--67, 2004.

\bibitem{li2005genre}
M.~Li and R.~Sleep, ``Genre classification via an {LZ78}-based string kernel,''
  in \emph{Proc.~6th Intern.~Conf.~Music Information Retrieval (ISMIR)}, 2005.

\bibitem{helen2007similarity}
M.~Hel{\'e}n and T.~Virtanen, ``A similarity measure for audio query by example
  based on perceptual coding and compression,'' in \emph{Proc.~10th
  Intern.~Conf.~Digital Audio Effects (DAFX)}, 2007.

\bibitem{ahonen2009measuring}
T.~E. Ahonen, ``Measuring harmonic similarity using {PPM}-based compression
  distance,'' in \emph{Proc.~Workshop Exploring Musical Information Spaces
  (WEMIS)}, 2009, pp. 52--55.

\bibitem{bello2011measuring}
J.~P. Bello, ``Measuring structural similarity in music,'' \emph{IEEE
  Trans.~Audio, Speech, Language Process.}, vol.~19, no.~7, pp. 2013--2025,
  2011.

\bibitem{bertin2011million}
T.~Bertin-Mahieux, D.~Ellis, B.~Whitman, and P.~Lamere, ``The million song
  dataset,'' in \emph{ISMIR 2011: Proceedings of the 12th International Society
  for Music Information Retrieval Conference, October 24-28, 2011, Miami,
  Florida}.\hskip 1em plus 0.5em minus 0.4em\relax University of Miami, 2011,
  pp. 591--596.

\bibitem{berenzweig2004large}
A.~Berenzweig, B.~Logan, D.~P.~W. Ellis, and B.~Whitman, ``A large-scale
  evaluation of acoustic and subjective music-similarity measures,''
  \emph{Computer Music Journal}, vol.~28, no.~2, pp. 63--76, 2004.

\bibitem{logan2001music}
B.~Logan and A.~Salomon, ``A music similarity function based on signal
  analysis,'' in \emph{Proc.~IEEE Intern.~Conf.~Multimedia and Expo.~(ICME)},
  2001, pp. 745--748.

\bibitem{aucouturier2002music}
J.~Aucouturier and F.~Pachet, ``Music similarity measures: What's the use?'' in
  \emph{Proc.~3rd Intern.~Conf.~Music Information Retrieval (ISMIR)}, 2002, pp.
  157--163.

\bibitem{aucouturier2007bag}
J.~Aucouturier, B.~Defreville, and F.~Pachet, ``The bag-of-frames approach to
  audio pattern recognition: A sufficient model for urban soundscapes but not
  for polyphonic music,'' \emph{Journal Acoustical Society of America}, vol.
  122, pp. 881--891, 2007.

\bibitem{fu2011music}
Z.~Fu, G.~Lu, K.~M. Ting, and D.~Zhang, ``Music classification via the
  bag-of-features approach,'' \emph{Pattern Recognition Letters}, vol.~32,
  no.~14, pp. 1768--1777, 2011.

\bibitem{marko2010audio}
M.~Hel{\'e}n and T.~Virtanen, ``Audio query by example using similarity
  measures between probability density functions of features,'' \emph{EURASIP
  Journal Audio, Speech, and Music Process.}, vol. 2010, 2010.

\bibitem{lerdahl1996generative}
F.~Lerdahl and R.~Jackendoff, \emph{A Generative Theory of Tonal Music}.\hskip
  1em plus 0.5em minus 0.4em\relax MIT Press, 1983.

\bibitem{casey2006importance}
M.~A. Casey and M.~Slaney, ``The importance of sequences in musical
  similarity,'' in \emph{Proc.~IEEE Intern.~Conf.~Acoustics, Speech and Signal
  Process.~(ICASSP)}, vol.~5, 2006.

\bibitem{paulus2010state}
J.~Paulus, M.~M{\"u}ller, and A.~Klapuri, ``State of the art report:
  Audio-based music structure analysis,'' in \emph{Proc.~11th Intern.~Society
  for Music Information Retrieval Conf.~(ISMIR)}, 2010, pp. 625--636.

\bibitem{fujishima1999realtime}
T.~Fujishima, ``Realtime chord recognition of musical sound: a system using
  common {L}isp music,'' in \emph{Proc.~Intern.~Computer Music Conf.~(ICMC)},
  1999, pp. 464--467.

\bibitem{bartsch2001catch}
M.~A. Bartsch and G.~H. Wakefield, ``To catch a chorus: Using chroma-based
  representations for audio thumbnailing,'' in \emph{IEEE Workshop Applications
  of Signal Process.~to Audio and Acoustics (WASPAA)}, 2001, pp. 15--18.

\bibitem{foote2000arthur}
J.~Foote, ``{ARTHUR}: Retrieving orchestral music by long-term structure,'' in
  \emph{Proc.~Intern.~Symp.~Music Information Retrieval (ISMIR)}, 2000.

\bibitem{gomez2006song}
E.~G{\'o}mez and P.~Herrera, ``The song remains the same: Identifying versions
  of the same piece using tonal descriptors,'' in \emph{Proc.~7th
  Intern.~Conf.~Music Information Retrieval (ISMIR)}, 2006.

\bibitem{serra2009cross}
J.~Serr{\`a}, X.~Serra, and R.~G. Andrzejak, ``Cross recurrence quantification
  for cover song identification,'' \emph{New Journal of Physics}, vol.~11,
  no.~9, p. 093017, 2009.

\bibitem{serra2012predictability}
J.~Serr{\`a}, H.~Kantz, X.~Serra, and R.~G. Andrzejak, ``Predictability of
  music descriptor time series and its application to cover song detection,''
  \emph{IEEE Trans.~Audio, Speech, Language Process.}, vol.~20, no.~2, pp.
  514--525, 2012.

\bibitem{ellis2007identifyingcover}
D.~P.~W. Ellis and G.~E. Poliner, ``Identifying `cover songs' with chroma
  features and dynamic programming beat tracking,'' in \emph{Proc.~IEEE
  Intern.~Conf.~Acoustics, Speech and Signal Process.~(ICASSP)}, vol.~4, 2007,
  pp. 1429--1432.

\bibitem{jensen2008chroma}
J.~H. Jensen, M.~G. Christensen, and S.~H. Jensen, ``A chroma-based
  tempo-insensitive distance measure for cover song identification using the
  2{D} autocorrelation function,'' in \emph{Music Information Retrieval
  Evaluation Exchange Task Audio Cover Song Identification}, 2008.

\bibitem{bertinlarge}
T.~Bertin-Mahieux and D.~P.~W. Ellis, ``Large-scale cover song recognition
  using the 2{D} {F}ourier transform magnitude,'' in \emph{Proc.~13th
  Intern.~Society for Music Information Retrieval Conf.~(ISMIR)}, 2012, pp.
  241--246.

\bibitem{tsai2005query}
W.~Tsai, H.~Yu, and H.~Wang, ``A query-by-example technique for retrieving
  cover versions of popular songs with similar melodies,'' in \emph{Proc.~6th
  Intern.~Conf.~Music Information Retrieval (ISMIR)}, 2005, pp. 183--190.

\bibitem{bello2007audio}
J.~P. Bello, ``Audio-based cover song retrieval using approximate chord
  sequences: testing shifts, gaps, swaps and beats,'' in \emph{Proc.~8th
  Intern.~Conf.~Music Information Retrieval (ISMIR)}, 2007, pp. 239--244.

\bibitem{lee2006identifying}
K.~Lee, ``Identifying cover songs from audio using harmonic representation,''
  in \emph{Music Information Retrieval Evaluation Exchange Task Audio Cover
  Song Identification}, 2006.

\bibitem{martin2012blast}
B.~Martin, D.~G. Brown, P.~Hanna, and P.~Ferraro, ``{BLAST} for audio sequences
  alignment: a fast scalable cover identification tool,'' in \emph{Proc.~13th
  Intern.~Society for Music Information Retrieval Conf.~(ISMIR)}, 2012.

\bibitem{ahonen2010combining}
T.~Ahonen, ``Combining chroma features for cover version identification,'' in
  \emph{Proc.~11th Intern.~Society for Music Information Retrieval
  Conf.~(ISMIR)}, 2010, pp. 165--170.

\bibitem{cleary1984data}
J.~Cleary and I.~Witten, ``Data compression using adaptive coding and partial
  string matching,'' \emph{IEEE Trans.~Commun.}, vol.~32, no.~4, pp. 396--402,
  1984.

\bibitem{ahonen2012compression}
T.~Ahonen, ``Compression-based clustering of chromagram data: New method and
  representations,'' in \emph{Proc.~9th Intern.~Symposium Computer Music
  Modeling and Retrieval}, 2012, pp. 474--481.

\bibitem{burrows1994block}
M.~Burrows and D.~J. Wheeler, ``A block-sorting lossless data compression
  algorithm,'' Digital Equipment Corporation, Tech. Rep., 1994.

\bibitem{tabus2012information}
I.~Tabus, V.~Tabus, and J.~Astola, ``Information theoretic methods for aligning
  audio signals using chromagram representations,'' in \emph{Proc.~5th
  Intern.~Symp.~Communications Control and Signal Process.~(ISCCSP)}, 2012, pp.
  1--4.

\bibitem{silva2013video}
D.~Silva, H.~Papadopoulos, G.~Batista, and D.~Ellis, ``A video
  compression-based approach to measure music structural similarity,'' in
  \emph{Proc.~14th Intern.~Society for Music Information Retrieval
  Conf.~(ISMIR)}, 2013, pp. 95--100.

\bibitem{casey2008analysis}
M.~Casey, C.~Rhodes, and M.~Slaney, ``Analysis of minimum distances in
  high-dimensional musical spaces,'' \emph{IEEE Trans.~Audio, Speech, and
  Language Process.}, vol.~16, no.~5, pp. 1015--1028, 2008.

\bibitem{slaney2008locality}
M.~Slaney and M.~Casey, ``Locality-sensitive hashing for finding nearest
  neighbors,'' \emph{IEEE Signal Process.~Magazine}, vol.~25, no.~2, pp.
  128--131, 2008.

\bibitem{bertin2011large}
T.~Bertin-Mahieux and D.~Ellis, ``Large-scale cover song recognition using
  hashed chroma landmarks,'' in \emph{IEEE Workshop Applications of Signal
  Process.~to Audio and Acoustics (WASPAA)}.\hskip 1em plus 0.5em minus
  0.4em\relax IEEE, 2011, pp. 117--120.

\bibitem{khadkevichlarge}
M.~Khadkevich and M.~Omologo, ``Large-scale cover song identification using
  chord profiles,'' in \emph{Proc.~14th Intern.~Society for Music Information
  Retrieval Conf.~(ISMIR)}, 2013, pp. 233--238.

\bibitem{schnitzer2009filter}
D.~Schnitzer, A.~Flexer, and G.~Widmer, ``A filter-and-refine indexing method
  for fast similarity search in millions of music tracks.'' in \emph{Proc.~10th
  Intern.~Society for Music Information Retrieval Conf.~(ISMIR)}, 2009, pp.
  537--542.

\bibitem{loewenstern1995dna}
D.~Loewenstern, H.~Hirsh, P.~Yianilos, and M.~Noordewier, ``{DNA} sequence
  classification using compression-based induction,'' Center for Discrete
  Mathematics and Theoretical Computer Science, Tech. Rep., 1995.

\bibitem{ziv1977universal}
J.~Ziv and A.~Lempel, ``A universal algorithm for sequential data
  compression,'' \emph{IEEE Trans.~Inf.~Theory}, vol.~23, no.~3, pp. 337--343,
  1977.

\bibitem{ziv1993measure}
J.~Ziv and N.~Merhav, ``A measure of relative entropy between individual
  sequences with application to universal classification,'' \emph{IEEE
  Trans.~Inf.~Theory}, vol.~39, no.~4, pp. 1270--1279, 1993.

\bibitem{benedetto2002language}
D.~Benedetto, E.~Caglioti, and V.~Loreto, ``Language trees and zipping,''
  \emph{Physical Review Letters}, vol.~88, no.~4, p. 48702, 2002.

\bibitem{cilibrasi2005clustering}
R.~Cilibrasi and P.~M.~B. Vit{\'a}nyi, ``Clustering by compression,''
  \emph{IEEE Trans.~Inf.~Theory}, vol.~51, no.~4, pp. 1523--1545, 2005.

\bibitem{sculley2006compression}
D.~Sculley and C.~E. Brodley, ``Compression and machine learning: A new
  perspective on feature space vectors,'' in \emph{Proc.~Data Compression
  Conf.~(DCC)}, 2006, pp. 332--341.

\bibitem{grunwald2004shannon}
P.~Gr{\"u}nwald and P.~M.~B. Vit{\'a}nyi, ``Shannon information and
  {K}olmogorov complexity,'' \emph{arXiv e-print cs/0410002}, 2004.

\bibitem{kaltchenko2004algorithms}
A.~Kaltchenko, ``Algorithms for estimating information distance with
  application to bioinformatics and linguistics,'' in \emph{Proc.~IEEE Canadian
  Conf.~Electrical and Computer Engineering (CCECE)}, vol.~4, 2004, pp.
  2255--2258.

\bibitem{feder1992universal}
M.~Feder, N.~Merhav, and M.~Gutman, ``Universal prediction of individual
  sequences,'' \emph{IEEE Trans.~Inf.~Theory}, vol.~38, no.~4, pp. 1258--1270,
  1992.

\bibitem{feder1994relations}
M.~Feder and N.~Merhav, ``Relations between entropy and error probability,''
  \emph{IEEE Trans.~Inf.~Theory}, vol.~40, no.~1, pp. 259--266, 1994.

\bibitem{begleiter2004prediction}
R.~Begleiter, R.~El-Yaniv, and G.~Yona, ``On prediction using variable order
  {M}arkov models,'' \emph{Journal of Artificial Intelligence Research},
  vol.~22, pp. 385--421, 2004.

\bibitem{takens1981detecting}
F.~Takens, ``Detecting strange attractors in turbulence,'' \emph{Dynamical
  systems and turbulence}, pp. 366--381, 1981.

\bibitem{gomez2006tonal}
E.~G{\'o}mez, ``Tonal description of music audio signals,'' Ph.D. dissertation,
  Universitat Pompeu Fabra, 2006.

\bibitem{foster2011causal}
P.~Foster, A.~Klapuri, and M.~D. Plumbley, ``Causal prediction of
  continuous-valued music features,'' in \emph{Proc.~12th Intern.~Society for
  Music Information Retrieval Conf.~(ISMIR)}, 2011, pp. 501--506.

\bibitem{ellis2006beat}
D.~P.~W. Ellis, ``Beat tracking with dynamic programming,'' in \emph{Music
  Information Retrieval Evaluation Exchange Tasks on Audio Tempo Extraction and
  Audio Beat Tracking}, 2006.

\bibitem{echonestfeatures}
T.~Jehan, ``Analyzer documentation,'' The Echo Nest, Tech. Rep., 2011.

\bibitem{ravuri2010cover}
S.~Ravuri and D.~P.~W. Ellis, ``Cover song detection: from high scores to
  general classification,'' in \emph{Proc.~IEEE Intern.~Conf.~Acoustics Speech
  and Signal Process.~(ICASSP)}, 2010, pp. 65--68.

\bibitem{osmalskyj2013efficient}
J.~Osmalskyj, S.~Pierard, M.~Van~Droogenbroeck, and J.~Embrechts, ``Efficient
  database pruning for large-scale cover song recognition,'' in
  \emph{Proc.~IEEE Intern.~Conf.~Acoustics, Speech and Signal
  Process.~(ICASSP)}.\hskip 1em plus 0.5em minus 0.4em\relax IEEE, 2013.

\bibitem{friedman1937use}
M.~Friedman, ``The use of ranks to avoid the assumption of normality implicit
  in the analysis of variance,'' \emph{Journal American Statistical
  Association}, vol.~32, no. 200, pp. 675--701, 1937.

\bibitem{tukey1973problem}
J.~W. Tukey, \emph{The Problem of Multiple Comparisons}.\hskip 1em plus 0.5em
  minus 0.4em\relax Princeton University, 1973.

\bibitem{kittler1998combining}
J.~Kittler, M.~Hatef, R.~P.~W. Duin, and J.~Matas, ``On combining
  classifiers,'' \emph{IEEE Trans.~Pattern Analysis and Machine Intelligence},
  vol.~20, no.~3, pp. 226--239, 1998.

\bibitem{shalizi2004blind}
C.~R. Shalizi and K.~L. Shalizi, ``Blind construction of optimal nonlinear
  recursive predictors for discrete sequences,'' in \emph{Proc.~20th
  Conf.~Uncertainty in Artificial Intelligence (UAI)}, 2004, pp. 504--511.

\bibitem{hochreiter1997long}
S.~Hochreiter and J.~Schmidhuber, ``Long short-term memory,'' \emph{Neural
  computation}, vol.~9, no.~8, pp. 1735--1780, 1997.

\end{thebibliography}

\end{document}